\documentclass{iopart}

\usepackage{color}
\usepackage{graphicx}

\def\gsim{\mathop {\vtop {\ialign {##\crcr 
$\hfil \displaystyle {>}\hfil $\crcr \noalign {\kern1pt \nointerlineskip } 
$\,\sim$ \crcr \noalign {\kern1pt}}}}\limits}
\def\lsim{\mathop {\vtop {\ialign {##\crcr 
$\hfil \displaystyle {<}\hfil $\crcr \noalign {\kern1pt \nointerlineskip } 
$\,\,\sim$ \crcr \noalign {\kern1pt}}}}\limits}

\def\ggs{\buildrel\textstyle > \over {\hbox{\raise0.2ex\hbox{$\sim$}}}}
\def\lls{\buildrel\textstyle < \over {\hbox{\raise0.2ex\hbox{$\sim$}}}}
\def\gsim{\, \lower0.75ex\hbox{$\ggs$}\, }
\def\lsim{\, \lower0.75ex\hbox{$\lls$}\, }
\newcommand{\beq}{\begin{equation}}
\newcommand{\eeq}{\end{equation}}

\newcommand{\beqa}{\begin{eqnarray}}
\newcommand{\eeqa}{\end{eqnarray}}
\newcommand{\non}{\nonumber}

\newcommand{\mib}[1]{\mbox{\boldmath${#1}$}}
\newcommand{\citen}[1]{\cite{#1}}

\def \e{{\rm e}}	
\def \Tr{{\rm Tr}}	

\def \Hl{H_{\lambda}}
\def \H0{H^{(0)}}

\def \ep{\varepsilon}	
\def \Q{\hat Q}

\def \phi{{\varphi}}

\begin{document}

\title{
Unconventional Non-Fermi Liquid Properties of Two-Channel Anderson Impurities System
}

\author{{Atsushi Tsuruta}$^{1}$ and {Kazumasa Miyake}$^{2}$ 
}
\address{
$^{1}$Division of Materials Physics, Department of Materials Engineering Science,
Graduate School of Engineering Science, Osaka University, Toyonaka, Osaka
560-8531, Japan\\
$^{2}$Center for Advanced High Magnetic Field Science, Osaka University, Toyonaka, 
Osaka 560-0043, Japan \\
}
\ead{tsuruta@mp.es.osaka-u.ac.jp}
\vspace{10pt}
\begin{indented}
\item[]April 2021
\end{indented}

\begin{abstract}
A theory for treating the unconventional non-Fermi liquid temperature dependence of 
physical quantities, such as the resistivity, in the Pr-based two-channel Anderson impurities 
system is developed. It is shown that their temperature dependences are essentially the same as 
those in the pure lattice system except for the case of extremely low concentration of Pr ions 
that is difficult to realize by the controlled experiments.  
This result is consistent with recent observations in diluted Pr-1-2-20 system  
Y$_{1-x}$Pr$_x$Ir$_2$Zn$_{20}$ ($x=0.024,\,0.044,\,0.085,$ and $0.44$) reported in Yamane {\it et al}. 
Phys. Rev. Lett. {\bf 121}, 077206 (2018), 
and is quite different from that in the case of single-channel 
Anderson impurities system in which the crossover between behaviors of the local Fermi liquid and heavy 
Fermi liquid occurs at around moderate concentration of impurities as observed in Ce-based heavy fermion system
La$_{1-x}$Ce$_x$Cu$_6$.
\end{abstract}

\section{Introduction}
In a past decade, non-Fermi liquid behaviors observed in the so-called 
Pr-1-2-20 compounds, Pr$T_2A_{20}$ ($T$=Ti, V; Rh, Ir and $A$=Al; Zn), have attracted much 
attention ~\cite{Sakai,Onimaru1,Onimaru1A,Onimaru4}. 
Namely, the temperature ($T$) dependence of the electrical resistivity 
follows $\sqrt{T}$-like behavior in a wide $T$ region.
The specific heat, $C(T)$, and the magnetic susceptibility, $\chi_m(T)$, 
increase in proportion to $({\rm const.}-\sqrt{T})$ toward $T_Q$, the transition temperature of quadrupolar ordering, as $T$ decreases below the Kondo temperature $T_{\rm K}$  
that is a fundamental energy scale characterizing the physics. 
Similar anomalies have been reported
in PrPb$_3$~\cite{Sato}.
The common aspect of these compounds is that 
the ground state of the crystalline-electrical field (CEF) of the localized 4f-electron  in Pr$^{3+}$ ion  
is the $\Gamma_3$ non-Kramers doublet in 4f$^2$-configuration, 
as verified by analyses of the specific heat and the magnetic moment, and inelastic neutron 
scattering experiments~\cite{Onimaru1A}.

Such a system with the $\Gamma_3$ CEF ground state in f$^2$-configuration is expected to exhibit anomalous behaviors associated with the two-channel Kondo 
effect~\cite{Nozieres,Cox1,Cox3,Cox2}.
Indeed, 
we have shown that the above mentioned anomalous properties can be understood in a unified fashion 
on the basis of the two-channel Anderson lattice model~\cite{Tsuruta6}, 
with the use of $1/N$ expansion method 
{\it \`a la} Nagoya group~\cite{Ono,Tsuruta4,Tsuruta5,Nishida} that makes it possible 
to take into account the strong correlation effect properly by satisfying a series of constrains 
among auxiliary  particles, slave fermions and bosons, in each order of $1/N$.  
In particular, the $T$ dependence of the resistivity $\rho(T)$ is given 
in the form~\cite{Tsuruta6}
\beqa
\rho(T)=\frac{aT}{T+bT_{0}}\left(1-\frac{1}{M^{2}}\right)+c\left(\frac{T}{T_{\rm K}}\right)^{2},
\label{rho_T}
\eeqa
where $M$ is the number of channel, and 
$T_{0}$ is the temperature characterizing the non-Fermi liquid state below which the 
$\sqrt{T}$-like dependence in the resistivity is observed in wide $T$ 
region although the resistivity follows the $T$-linear dependence in the low temperature limit 
$T \ll T_{0}$.  
Note that this $T_{0}$ is a new temperature scale characterizing the two-channel lattice Anderson or Kondo system which should be distinguished from the so-called Kondo temperature $T_{\rm K}$ for the single-channel lattice Anderson or Kondo system.
The coefficients $a$ and $c$ in Eq.\ (\ref{rho_T}) are constants with dimension of 
the resistivity, and 
depend on the materials parameters 
characterizing the system, such as the strength of the c-f hybridization.     
On the other hand, $b$ is approximately given by $b\simeq0.67$ as discussed in Appendix A.
It is crucial that the non-Fermi liquid term scaled by $T_{0}$ in 
Eq.\ (\ref{rho_T}) exists only in the case of multi-channel with $M\ge 2$.   
This scaling behavior has really been observed in a series of 
Pr-1-2-20 compounds with different $T_{0}$s depending on the pressure and the magnetic 
field~\cite{Onimaru3,Yoshida_Izawa}.
The relationship between the scaling form given by
Eq.\ (\ref{rho_T}) and the scaling behaviors observed in Refs. \citen{Onimaru3,Yoshida_Izawa} is 
discussed in Appendix A.

On the other hand, it has been reported recently in diluted systems
Y$_{1-x}$Pr$_x$Ir$_2$Zn$_{20}$ ($x=0.024,\,0.044,\,0.085,\,0.44$) \cite{Yamane}
that the $T$ dependence in the resistivity $\rho(T)$ follows essentially the same as that of the lattice system, i.e., $x=1$.
This contradicts the theoretical result for the two-channel ($M=2$) 
impurity Kondo effect~\cite{Ludvig} 
according to which $\rho(T)\propto ({\rm const.}\pm\sqrt{T})$ depending whether the exchange 
interaction is in the strong coupling region (+) or weak coupling region (--). 
If the system is located in the strong-coupling region, observed $T$-dependence could be understood 
as the two-channel impurity Kondo effect.  However, since it is rather hard 
to expect that these systems are located in the strong coupling region, this conflicting behavior 
cannot be understood as the two-channel impurity Kondo effect, offering theorists  
a big challenge.  

In this paper, we show theoretically that the two-channel Anderson impurities system as 
Y$_{1-x}$Pr$_x$Ir$_2$Zn$_{20}$ ($10^{-6}\lsim x\ll 1$) exhibits essentially the same non-Fermi liquid behaviors as 
the pure system with $x=1$ unless $x$ is extremely small less than $x^{\rm cr}\sim 10^{-6}$ 
for a reasonable set of parameters. The organization of the present paper is as follows. 
In Sect.\ 2, the model for the two-channel Anderson impurities system is introduced, and 
an outline of the $1/N$-expansion method for taking the strong correlation effect 
is recapitulated. 
In Sect.\ 3, a new formalism of treating the effect of random distribution of dilute Pr ions is 
proposed.  
In Sect.\ 4, on the basis of the formalism developed in Sect.\ 3, the $T$ dependence 
of the resistivity in the two-channel Anderson impurities system is derived. 
In Sect.\ 5, the $T$ dependence of physical quantities is summarized. 
In Sect.\ 6, the critical impurity concentration $c_{\rm imp}^{\rm cr} (x^{\rm cr})$ below which the resistivity 
shows the $T$ dependence of the single impurity model is discussed, showing that 
$c_{\rm imp}^{\rm cr}$ for the two-channel model is extremely small 
of ${\cal O}(10^{-6})$ and not reached by controlled 
experiments, while that for the single-channel model is only moderately smaller than 1 in consistent 
with observation in Ce-based impurity heavy fermion systems such as Ce$_x$La$_{1-x}$Cu$_6$
\cite{Sumiyama,Onuki}.

\section{Model Hamiltonian and Formulation}

\begin{figure}
\includegraphics[clip,width=10cm]{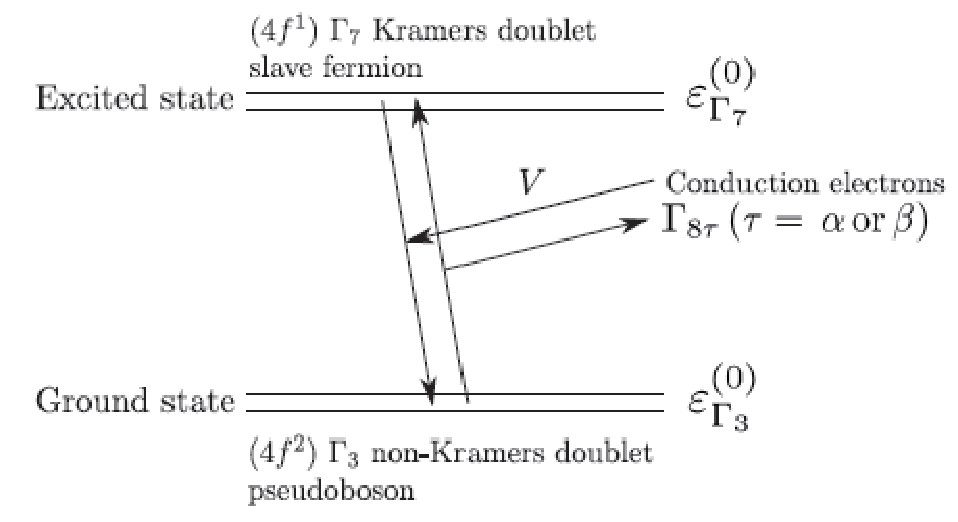}
\caption{Levels scheme of f-electrons and hybridization path with conduction electrons of model Hamiltonian [Eq.\ (\ref{H})].}
\label{Fig:level}
\end{figure}

A canonical model for describing Pr-1-2-20 compounds, in which the CEF ground state 
of Pr$^{3+}$ ion in 4f$^{2}$ configuration 
is the $\Gamma_{3}$ non-Kramers doublet and hybridizing with conduction electrons with $\Gamma_8$ 
symmetry leaving the 4f$^1$ $\Gamma_7$ Kramers doublet (as shown in Fig.\ \ref{Fig:level}), 
is given by the two-channel Anderson lattice model discussed in Refs.~\citen{Tsuruta4,Tsuruta5,Tsuruta6}: 
\beqa
&&
H_{\rm 2cAL}=\sum_{\sigma=1}^{M}\sum_{\tau_1,\tau_2=1}^{N} \sum_{\mib{k}} \ep_{\mib{k}\tau_1\tau_2} c^+_{\mib{k}\tau_1\sigma} c_{\mib{k}\tau_2\sigma}
\non\\
&&\hspace{2cm} 
+\sum_i\sum_{\tau=1}^{N} \ep^{(0)}_{\Gamma_3} b^+_{i\tau} b_{i\tau}
+\sum_i\sum_{\sigma=1}^{M} \ep^{(0)}_{\Gamma_7} f^+_{i\sigma} f_{i\sigma}
\non\\
&&\hspace{2cm} 
+\frac{1}{\sqrt{N_L}}\sum_{\sigma=1}^M\sum_{\tau=1}^N\sum_{i, \mib{k}} \left (V c^+_{\mib{k}\tau\bar{\sigma}} b_{i \tau} f^+_{i\sigma}e^{-{\rm i} \mib{k}\cdot\mib{R}_i} + {\rm h.c.} \right ),
\label{H}
\eeqa
where $c_{\mib{k}\tau\sigma}$ is the annihilation operators of a conduction electron with wave vector 
$\mib{k}$ and dispersion $\varepsilon_{\mib{k}\tau_{1}\tau_{2}}$, 
and spin-orbital component $\sigma$ (with $\bar{\sigma}$ being the opposite component 
of $\sigma$) specifying the CEF ground state of $\Gamma_{8}$ in the 4f$^{1}$ configuration 
($M=2$), 
$b_{i\tau}$  is that of the pseudo boson representing the $i$-th localized f$^2$ state 
of energy $\ep^{(0)}_{\Gamma_3}$ 
with the symmetry of 
$\Gamma_3$ with quadrupole moment $\tau$ ($N=2$), 
$f_{i\sigma}$ is that of the pseudo fermion representing the $i$-th localized f$^1$ state 
of energy $\ep^{(0)}_{\Gamma_7}$ 
with spin-orbital momentum $\sigma$ specifying the CEF ground state of $\Gamma_{7}$, and  
$V$ represents the hybridization transforming from 
the f$^2$-state with $\tau$ at $i$-th site to the composite state of f$^1$ with $\sigma$ at the same site 
and that described by $c_{\mib{k}\tau\bar{\sigma}}$, and vice versa, as shown in Fig.\ \ref{Fig:level}. 
Note that $N_L$ is the total number of lattice sites for conduction electrons while the f-electrons 
occupy only the sites of Pr ions that is dilutely distributed on the $N_{\rm L}$ 
lattice sites, and $N=2$ and $M=2$ 
stand for components of the spin-orbital degeneracy and that of the quadrupole moment, respectively. 
Hereafter, we discard $\ep_{\mib{k}\tau{\bar \tau}}$, which is non zero in general, 
because there occurs no qualitative difference from the case $\ep_{\mib{k}\tau{\bar \tau}}\not=0$ 
as described in Ref.~\citen{Tsuruta6}.

To guarantee that the transformed model [Eq. (\ref{H})] describes the physical process shown in 
Fig.\ \ref{Fig:level}, the Hamiltonian [Eq. (\ref{H})] must be treated within the subspace 
where the local constraint
\beq
\Q_{i} = \sum_{\tau}b^+_{i \tau} b_{i \tau} + \sum_{\sigma} f^+_{i\sigma} f_{i\sigma} = 1, \label{Q}
\eeq
is fulfilled.
To calculate physical quantities within the subspace restricted by the local constraint [Eq. (\ref{Q})], we evaluate the expectation value $\langle\hat{A}\rangle$ of a physical quantity $\hat{A}$ such that~\cite{Coleman,Jin}  
\beqa
\left\langle \hat{A} \right\rangle = \lim_{\{\lambda_i\} \rightarrow \infty}
		\left( {\left\langle \hat{A} \Pi_i \Q_{i} \right\rangle}_\lambda / 
		{\left\langle \Pi_i \Q_{i} \right\rangle}_\lambda \right), 
			 \label{expectation value}
\eeqa
where
\beqa
\langle \hat{A} \Pi_i \Q_{i} \rangle _\lambda \equiv
		\Tr[\e^{-\beta \Hl} \hat{A} \Pi_i \Q_{i}]/Z_\lambda,
\eeqa
 with
\beqa 
&&Z_\lambda \equiv \Tr[\e^{-\beta \Hl}],\\
&&\Hl \equiv H + \sum_i \lambda_i \Q_{i}.
\eeqa
In order to calculate the average [Eq. (\ref{expectation value})] explicitly, we employ the perturbation expansion 
in the power of $1/N$ following the rules as
\beqa
\frac1{N_L}\sum_{\tau}\sum_{\mib{k}}1=O\left[ (1/N)^0  \right],
\eeqa
and
\beqa
\frac1{N_L}\sum_{\mib{k}}1=O\left( 1/N  \right).\label{O1N}
\eeqa
In Refs.~\citen{Ono,Nishida}, one can see the validity of this rule of power counting in $1/N$ and its physical meaning behind it.
For explicit calculations in this paper, we set $N=2$, which may not lose the generality because we do not use the condition $1/N\ll 1$ explicitly.

\section{Average over Random Distribution of Pr Ions}

A basic idea is that 
4f electrons at Pr sites acquire the wave 
vector ${\bf p}$ dependence through the average process 
over the random distribution of Pr ions, which in turn gives rise to two contributions to the scattering process of the 
conduction electrons, i.e., a single site effect of localized f-electrons and the lattice effect 
due to the wave-number dependent collective quadrupole 
fluctuations~\cite{Miyake_Watanabe}.
 
\begin{figure}[h]
\begin{center}
\rotatebox{0}{\includegraphics[width=0.8\linewidth]{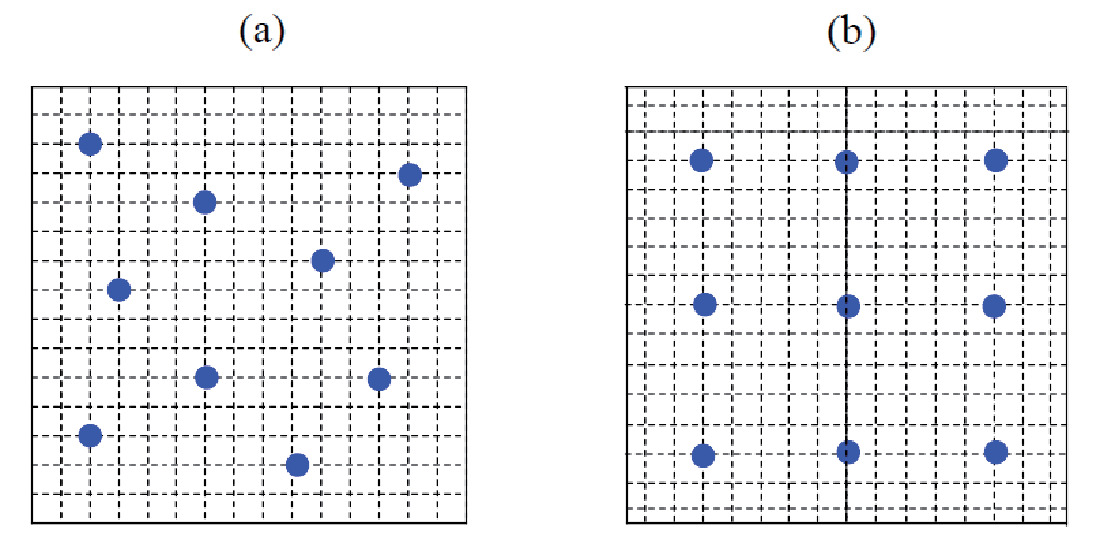}}
\caption{(a) Square lattice version of original system where the sites occupied by Pr ions are shown 
by filled circle. (b) {\it Virtual} system obtained after average over random distribution of Pr sites. 
Hereafter, conduction electrons are assumed to be described by wave vector ${\bf p}$ in both systems.  
}
\label{Fig:Replica}
\end{center}
\end{figure}

To estimate the effect of scattering due to the random distribution of Pr ions, 
we have to take an average over their random distribution.  
Before taking the average, the one-particle Green function of f-electrons depends on the two positions 
as $G_{\rm f}({\bf r}_{i}, {\bf r}_{j}; {\rm i}\varepsilon_n)$,
where $i\varepsilon_n$ is the fermionic Matsubara frequency,
but it becomes a function of the relative coordinate $({\bf r}_{i}-{\bf r}_{j})$ after taking 
the average over the random distribution of Pr ions
{
so that the Green function acquires the wave vector
representation in general.
This kind of procedure has been established in the case where the conduction electrons are influenced by the random distribution of impurities giving random potential,
as discussed, e.g., in the textbook \citen{AGD}.
However, this acquirement of the diagonal wave vector dependence or the recovery of translational symmetry is only due to a general property of the random average.
Namely, this is {\it not} an approximation, but is based on the principle or "{\it Ansatz}" of physical statistics. In this sense, it is virtually rigorous.
In fact, the uniformity of the distribution of Pr impurities was confirmed experimentally with good accuracy at least in the case of Y$_{1-x}$Pr$_x$Ir$_2$Zn$_{20}$ \citen{Yamane,Comment_Onimaru}.
Therefore, }by taking this average, the wave vector dependent Green function becomes well defined as
{
\begin{eqnarray}
{\bar G}_{\rm {\tilde f}}({\bf p},{\rm i}\varepsilon_n)
=\sum_{(i-j)}e^{-{\rm i}{\bf p}\cdot({\bf r}_{i}-{\bf r}_{j})}
	\langle\langle G_{\rm f}({\bf r}_{i},{\bf r}_{j}; {\rm i}\varepsilon_n)\rangle\rangle, \label{eq. gf}
\end{eqnarray}
{
where $\langle\langle\cdots\rangle\rangle$ denotes the average over the random distribution of Pr ions.
{
By the inverse Fourier transformation, we can introduce the real space {{\it image}} behind the expression [Eq. (\ref{eq. gf})] as shown in Fig. \ref{Fig:Replica}(b) in which Pr ions are distributed regularly in a certain plane, e.g. $xy$ plane, on a {\it virtual} lattice assumed to be simple cubic for simplicity.
However, there exists no direct (or one-to-one) correspondence between positions of lattice points of the original and {\it virtual} lattice because the origin of the {\it virtual} lattice cannot be fixed by the inverse Fourier transformation that specifies only the dependence on the relative coordinate $({\bf r}_i-{\bf r}_j)$.
This is consistent with the fact that the average number of Pr ions at the original lattice point becomes uniform after the average over the random distributions of Pr ions.
}

As discussed below, with the use of this {\it virtual lattice} system, 
the temperature dependence of the damping rate of quasiparticles, consisted of Pr impurities,  can be estimated relying on the theory for periodic lattice system developed in  Ref. \cite{Tsuruta6} which explains quite well existing experimental facts in a unified way.
As a result, transport properties observed in diluted Pr-1-2-20 systems can be understood consistently.

Since the mean distance between localized 4f electrons is given by $(N_{\rm L}/N_{\rm f})^{1/3}a$, with $a$ and $N_{\rm f}$ being the lattice constant of the original lattice and the total number of localized 4f electrons of Pr ions, we can introduce the real space {{\it image}} behind the expression [Eq.(10)] as given by Fig. 2(b) in which Pr ions are distributed regularly on a {\it virtual} lattice.
Namely, and expression of the wave vector ${\bf p}$ is given by
\begin{eqnarray}
	{\bf p}=\frac{2\pi}{L}(n_x, n_y, n_z)\ \ 
\end{eqnarray}
where the integer $n_{\alpha}$ ($\alpha=x,\,y,\,z$) is restricted as $0\le|n_{\alpha}|\le (N_{\rm L}/{N_{\rm f}})^{1/3}/2$, $L$ is the length of one side of the cubic material and $N_{\rm f}$ is the number of Pr ions in the original lattice system.
}
However, we have not assumed that the Pr ions form a periodic lattice in the {{\it original}} lattice 
in Fig. 2(a).
{
Therefore, of course, this periodicity in the {{\it virtual}} lattice cannot be observed by diffraction methods, X-ray and/or neutron scattering, in the {{\it real}} crystal.
}
}
Hereafter, ${\bf p}$'s are used for the wave vectors in the {{\it virtual}} lattice obtained after 
the random average over the positions of Pr ions, and are distinguished from 
wave vectors ${\bf k}$'s in the original lattice shown in Fig.\ \ref{Fig:Replica}(a). 

On the other hand, conduction electrons described by wave vector ${\bf p}$ are essentially 
unaltered by the effect of scattering by Pr impurities  except for some broadening of the 
dispersion due to impurities scattering~\cite{AGD}.
Namely, for example, the density of states (DOS) 
of conduction electrons at the Fermi level are essentially unaltered.   
However, since the size of the Brillouin zone (BZ) of the {\it virtual} lattice is shortened 
by a factor $c_{\rm imp}^{1/3}$, 
the band of conduction electrons splits into multibands in the shortened and 
reduced BZ~{\cite{Comment_band_splitting}}.
However, hereafter for simplicity of presentation, we use an extended zone scheme for conduction electrons. 

Thus, we can discuss physical properties of the 2-channel Anderson lattice system 
on the {\it virtual} periodic lattice described by the {\it virtual} Hamiltonian for the 
{\it virtual} lattice shown in Fig.\ \ref{Fig:Replica}(b). 
The {\it virtual} Hamiltonian $H^{\rm v}_{\rm 2cAL}$ is given explicitly as follows:
\begin{eqnarray}
& &
H^{\rm v}_{\rm 2cAL}\equiv
\sum_{\sigma=1}^{N}\sum_{\tau_1,\tau_2=1}^{M} \sum_{{
	a},{\bf p}} \ep_{{
		a}{\bf p}\tau_1\tau_2} {\tilde c}^+_{{
			a}{\bf p}\tau_1\sigma} {\tilde c}_{{
				a}{\bf p}\tau_2\sigma}
	\non\\
& &
\qquad\qquad
+\sum_{\tilde i}\sum_{\tau=1}^{M} \ep^{(0)}_{\Gamma_3} {\tilde b}^+_{{\tilde i}\tau} {\tilde b}_{{\tilde i}\tau}
+\sum_{\tilde i}\sum_{\sigma=1}^{N} \ep^{(0)}_{\Gamma_7} 
{\tilde f}^+_{{\tilde i}\sigma} {\tilde f}_{{\tilde i}\sigma}
	\non\\
& &
\qquad\qquad
+\frac{1}{\sqrt{N_{\rm L}}}\sum_{\sigma=1}^N\sum_{\tau=1}^M\sum_{{\tilde i}, {
	a}, {\bf p}} 
\left (V {\tilde c}^+_{{
	a}{\bf p}\tau\bar{\sigma}} {\tilde b}_{{\tilde i} \tau}
{\tilde f}^+_{{\tilde i}\sigma}e^{-{\rm i} {\bf p}\cdot\mib{R}_{\tilde i}} 
+ {\rm h.c.} \right )
	\non\\
& &
\qquad\qquad
+\sum_{{\tilde i},\sigma}u_{\rm imp}P_{\tilde i}\,n_{{\tilde i}\sigma}^{{\tilde {\rm c}}}
\label{eq:Replica}
\end{eqnarray}
where 
$\tilde{c}_{a{\bf p}\tau\sigma}$ and $\tilde{f}_{{\tilde i}\sigma}$ are the annihilation operators 
of conduction electrons {
	at band $a$} and f electrons at ${\tilde i}$-th site in the {\it virtual} periodic lattice, respectively, 
and $n_{{\tilde i}\sigma}^{\tilde {\rm c}}$ 
is the numbers of conduction 
at ${\tilde i}$-th site (${\tilde i}=1,2, \cdots,N_{\rm f}$). 
Note that the factor $1/\sqrt{N_{\rm L}}$ in the 4th term of Eq.\ (\ref{eq:Replica}) 
reflects the fact that 
the f electrons are located on the periodic {\it virtual} lattice points of $N_{\rm f}$ while the conduction electrons 
are hoping among the original lattice points of $N_{\rm L}$ as discussed in Appendix B. 
The random variables $P_{\tilde i}$ in the last term of Eq.\ (\ref{eq:Replica}) represent 
the distribution of the strength of impurity scattering and arises from the random distribution of 
Pr ions on the original lattice shown in Fig.\ \ref{Fig:Replica}(a).  
With the use of this {\it virtual} Hamiltonian,  
the theoretical framework for discussing the two-channel Anderson lattice system can be applied 
as it stands except the effect of impurities scattering on conduction electrons~\cite{Tsuruta6}. 


{
In PrIr$_2$Zn$_{20}$, Zn atoms form icosahedral cages with cubic ($T_d$) symmetry around Pr atoms.
Even in the case of dilute Pr atoms in Pr$_{x}$Y$_{1-x}$Ir$_2$Zn$_{20}$ ($x\ll1$), the symmetry of the 4f electrons contained in the Pr atom is basically determined by the local Zn icosahedral cage around the Pr atom.
Therefore, the ground state of the CEF of the f-electron in the Pr atom is considered to remain basically as $\Gamma_3$,
while the cage containing Pr atoms may be slightly distorted due to the effect of the other cages containing Pr atoms that are dilute and randomly located around the Pr atom in attention.
However, the distortion is small enough
{
	because the cage is rigid
and $x$ is small enough in the system in question},
so that the effect is negligible in the zeroth-order approximation.
As circumstantial evidence, the $T$ dependence of the electrical resistivity in the Pr dilute system reported in Ref.~\citen{Yamane} shows basically the same non-Fermi liquid $T$ dependence as that of the system with $x=1$.
This is also consistent with the following experimental fact.
Indeed, the $T$ dependence of ultrasound dispersion in the same system shows that the width of the energy splitting of $\Gamma_3$ states, $\Delta_{\Gamma_3}$, is about 0.048K~\cite{Yanagisawa},
which is sufficiently lower than the characteristic temperature scale characterising the non-Fermi liquid state: $T_0=0.3$K~\cite{Yamane},
above which the $T$ dependence of the electrical resistivity deviates from $\sqrt{T}$ behavior.
It is obvious that if $\Delta_{\Gamma_3}$ is larger than $T_0$, the non-Fermi liquid temperature dependence disappears.
On this observation, we consider that $\Delta_{\Gamma_3}$ is small enough, giving no essential effect.
In this sense, we have assumed that $\Delta_{\Gamma_3}=0$.
}
{
We consider that this assumption is correct at $x\sim1$ and $x\ll1$ where non-Fermi liquid behaviors have been observed in experiments~\cite{Yamane}, while in the intermediate region we can not ignore the effect of breaking the cubic symmetry.
}

{
Concluding this section, we compare our theoretical framework, which takes into account the effect of the impurity scattering on the conduction electrons from randomly distributed f-electrons, with three traditional methods for 
{alloy systems} discussed in Ref.~\citen{Ziman}.

(1) Our method of introducing the {\it virtual lattice} proposed in the present paper is different from the ``virtual crystal approximation (VCA)'', which is the method used for a homogeneous model in which the parameters contained in the Hamiltonian are averaged by concentration of alloy.
For example, the VCA cannot describe the existence of the hybridization gap caused by the c-f hybridization when applied to the non-interacting Anderson model.
In the model that assumes a simple virtual crystal without f-f interaction,
it is impossible to discuss how the effect of the local correlation between f-electrons influences on the $T$ dependence of the electrical resistivity.
The {\it virtual lattice} {in the present paper} should be distinguished from VCA.

(2) Our method shares in part a common aspect with the  ``averaged t-matrix approximation (ATA)'', which is a good approximation in the dilute region of f-electrons.
For example, the problem associated with the hybridization gap mentioned above in (1) is cleared as far as in the dilute case of f-electrons where the direct f-f hopping is safely neglected.
However, we need to extend it further, to capture the effect of the local correlation between f-electrons and the scattering effect by randomly distributed f-electrons that is the origin of the resistivity.
In that sense, the method proposed in the present paper can be regarded as one of reasonable extensions of ATA.

(3) There is no restriction on the concentration of f-electrons, $c_{\rm imp}$, in the framework of the present paper,
in contrast to the ``coherent potential approximation (CPA)'', in which the c-f hybridization gap disappears where the direct f-f hopping is neglected in the dilute region as mentioned above in (2).
Indeed, the theoretical result, obtained by the CPA with the dynamical mean field theory (DMFT), does not seem to reproduce the Nordheim rule $\rho\propto x(1-x)$ observed in Ce$_x$La$_{1-x}$Cu$_6$~\cite{Sumiyama} in the dilute case $x=c_{\rm imp}<0.3$~\cite{Mutou}.
}
\section{ Dual Nature of Resistivity}
The original one-particle Green function $G_{\rm c}({\bf r}_{i},{\bf r}_{j};{\rm i}\varepsilon_{n})$ of conduction 
electrons satisfies the Dyson equation as 
\begin{eqnarray}
& &
G_{\rm c}({\bf r}_{i},{\bf r}_{j};{\rm i}\varepsilon_{n})=
G^{(0)}_{\rm c}({\bf r}_{i},{\bf r}_{j};{\rm i}\varepsilon_{n})
\nonumber 
\\
& &
\qquad\qquad
+\sum_{k,\ell}G^{(0)}_{\rm c}({\bf r}_{i},{\bf r}_{k};{\rm i}\varepsilon_{n})
\Sigma_{\rm c}({\bf r}_{k},{\bf r}_{\ell};{\rm i}\varepsilon_{n})
G_{\rm c}({\bf r}_{\ell},{\bf r}_{j};{\rm i}\varepsilon_{n}),
\label{Green_c}
\end{eqnarray}
where $G^{(0)}_{\rm c}({\bf r}_{i},{\bf r}_{j};{\rm i}\varepsilon_{n})$ is the non-interacting Green function of 
conduction electrons, and $\Sigma_{\rm c}({\bf r}_{k},{\bf r}_{\ell};{\rm i}\varepsilon_{n})$ is the self-energy 
arising from the hybridization between the f-electrons at Pr sites. 
Here, we have abbreviated the suffices $\sigma$ and $\tau$ specifying the spin-orbital and quadrupole 
degrees of freedom for the concise presentation. 
After taking  the average over 
the random distribution of Pr ions, both the Green function and the self-energy depend only on the 
relative coordinate $({\bf r}_{i}-{\bf r}_{j})$'s and acquire the wave vector representation.  
Namely, the Green function ${\bar G}_{\rm {\tilde c}}({\bf p},{\rm i}\varepsilon_{n})\equiv
\sum_{(i-j)}e^{-{\rm i}{\bf p}\cdot({\bf r}_{i}-{\bf r}_{j})}
\langle\langle G_{\rm c}({\bf r}_{i},{\bf r}_{j};{\rm i}\varepsilon_{n})\rangle\rangle$ satisfies the 
Dyson equation as 
\begin{eqnarray}
& &
{\bar G}_{\rm {\tilde c}}({\bf p},{\rm i}\varepsilon_{n})=
G^{(0)}_{\rm c}({\bf p},{\rm i}\varepsilon_{n})
+G^{(0)}_{\rm c}({\bf p},{\rm i}\varepsilon_{n}){\bar \Sigma}_{\rm {\tilde c}}({\bf p},{\rm i}\varepsilon_{n})
{\bar G}_{\rm {\tilde c}}({\bf p},{\rm i}\varepsilon_{n}),
\label{Green_c_p}
\end{eqnarray}
where $G^{(0)}_{\rm c}({\bf p},{\rm i}\varepsilon_{n})$ is the Green function of free band electrons 
on the original lattice, and the self-energy ${\bar \Sigma}_{\rm {\tilde c}}({\bf p},{\rm i}\varepsilon_{n})$
 is defined by \begin{eqnarray}
& &
{\bar \Sigma}_{\rm {\tilde c}}({\bf p},{\rm i}\varepsilon_{n})\equiv
\sum_{(i-j)}e^{-{\rm i}{\bf p}\cdot({\bf r}_{i}-{\bf r}_{j})}
\langle\langle \Sigma_{\rm c}({\bf r}_{i},{\bf r}_{j};{\rm i}\varepsilon_{n})\rangle\rangle.
\label{Sigma_c_p}
\end{eqnarray}

The $T$ dependence of the resistivity is essentially given by the imaginary part of 
the retarded function of ${\bar \Sigma}_{\tilde c}({\bf p},{\rm i}\varepsilon_{n})$ [Eq.\ (\ref{Sigma_c_p})], in which 
$\langle\langle \Sigma_{\rm c}({\bf r}_{i},{\bf r}_{j};{\rm i}\varepsilon_{n})\rangle\rangle$ consists of 
two parts as 
\begin{eqnarray}
& &
\langle\langle \Sigma_{\rm c}({\bf r}_{i},{\bf r}_{j};{\rm i}\varepsilon_{n})\rangle\rangle
=-\frac{{\rm i}}{2\tau_{\rm imp}}{\rm sgn}(\varepsilon_{n})\delta_{ij}
+\langle\langle \Delta\Sigma_{\rm c}({\bf r}_{i},{\bf r}_{j};{\rm i}\varepsilon_{n})\rangle\rangle,
\label{Sigma_c_p:2}
\end{eqnarray}
where ${\rm i}/2\tau_{\rm imp}$ represents the damping effect 
arising from independent static scattering by localized f electrons at site ${\bf r}_{i}$, 
and gives temperature- and energy-independent term to the resistivity 
that is proportional to the 
concentration $c_{\rm imp}$ of Pr ions on the original lattice shown in Fig.\ \ref{Fig:Replica}(a).  
The second part of Eq. (\ref{Sigma_c_p:2}) arises from the scattering 
in the two-channel Anderson lattice system described by the {\it virtual} Hamiltonian 
[Eq.\ (\ref{eq:Replica})].  
Namely, the Green function of conduction electrons 
${\bar G}_{\rm {\tilde c}}({\bf p},{\rm i}\varepsilon_{n})$ has the following structure: 
\begin{eqnarray}
& &
{\bar G}_{\rm {\tilde c}}({\bf p},{\rm i}\varepsilon_{n})=
\left[
{\rm i}\varepsilon_{n}-\xi_{\bf p}+\frac{{\rm i}}{2\tau_{\rm imp}}{\rm sgn}(\varepsilon_{n})
-
{
{\tilde V}
}
^{2}{\bar G}_{\rm {\tilde f}}({\bf p},{\rm i}\varepsilon_{n})
\right]^{-1},
\label{Green_c_p:2}
\end{eqnarray}
where  ${\tilde V}$ is the averaged c-f hybridization 
${\tilde V}=\sqrt{c_{\rm imp}}V$ as discussed in Appendix B (see Eq. (\ref{eq:A4})), 
and
${\bar G}_{\rm {\tilde f}}({\bf p},{\rm i}\varepsilon_{n})$ is the f-electron Green function 
given by the {\it virtual} Hamiltonian [Eq.\ (\ref{eq:Replica})] so that it is influenced also by the 
damping effect of the conduction electrons as discussed below.   

From the structure of the Green function of conduction electron [Eq.\ (\ref{Green_c_p:2})], one can see 
that there exist two contributions to the resistivity. 
One arises from the renormalized impurity scattering of conduction electrons.  
Another one arises from the scattering in 
the two-channel Anderson lattice system described by the {\it virtual} Hamiltonian [Eq.\ (\ref{eq:Replica})]. 
The latter is expected to have the same temperature dependence of the resistivity as 
that of the lattice system except for the residual part at $T=0$, i.e., 
\begin{eqnarray}
& &
\rho_{\rm lattice}(T)-\rho_{0}^{*}\simeq\frac{aT}{T+bT_{0}}
\left(1-\frac{1}{M^{2}}\right)+c\left(\frac{T}{T_{\rm K}}\right)^{2},
\label{Rho_lattice}
\end{eqnarray}
which is essentially the same as Eq.\ (\ref{rho_T}) 
as shown below because it is essentially independent of the impurity scattering rate 
$1/2\tau_{\rm imp}$ of conduction electrons. The residual resistivity 
$\rho_{0}^{*}$ represents the renormalized resistivity arising from the renormalized impurities
scattering rate 
$1/2{\tilde \tau}_{\rm imp}\equiv1/2\tau_{\rm imp}
+[-{\rm Im}\langle\langle\Sigma_{\rm c}^{\rm R}(k_{\rm F}^{*},T=0)\rangle\rangle]$, 
where $\langle\langle\Sigma_{\rm c}^{\rm R}(k_{\rm F}^{*},T=0)\rangle\rangle$ is the imaginary part of the 
self-energy of conduction electrons in the {\it virtual} system described by the Hamiltonian 
[Eq.\ (\ref{eq:Replica})]. 

\begin{figure}[h]
\begin{center}
\rotatebox{0}{\includegraphics[width=0.7\linewidth]{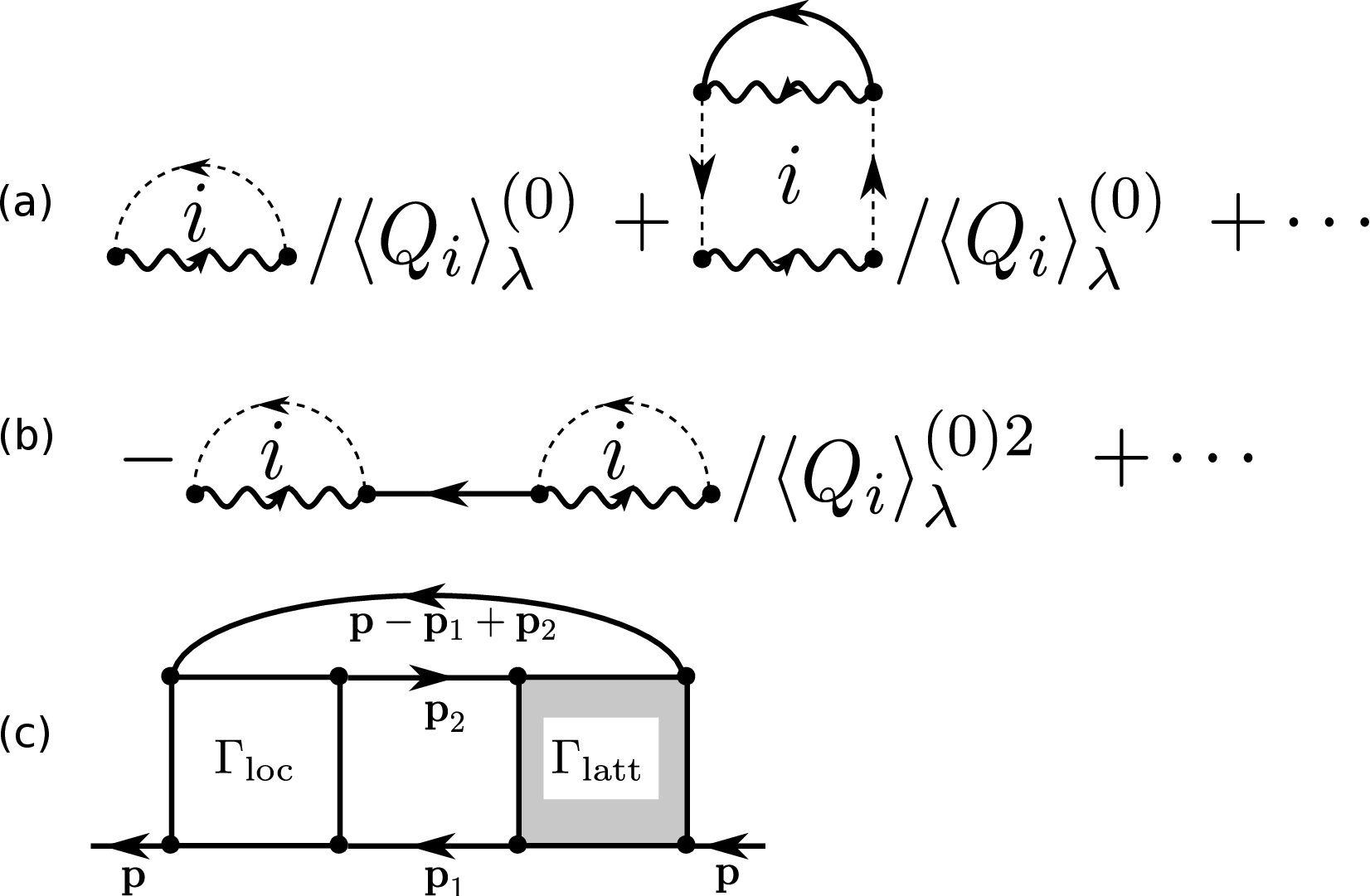}}
\caption{
Self-energies of conduction electrons in the {\it virtual} system shown in 
Fig.\ \ref{Fig:Replica}(b). 
Solid lines, dotted lines, and wavy lines represent the Green functions of conduction electrons, 
pseudo bosons, and slave fermions, respectively, and filled circles are the c-f hybridization 
${\tilde V}=\sqrt{c_{\rm imp}}V$.    
(a) Self-energy for the impurity model corresponding to the {\it virtual} model [Eq.(\ref{eq:Replica})]. 
The first and second terms represent the effect of ${\cal O}[(1/N)^{0}]$ and ${\cal O}[(1/N)^{1}]$, 
respectively, 
and the explicit form of higher order terms (dotted parts) are given 
by Eq.\ (4.6) in Ref.~\citen{Tsuruta2} representing the contribution of
Fig.\ 4 in Ref.~\citen{Tsuruta2}. 
(b) Self-energy for the local effect arising from the $d$-infinity {\it virtual} model [Eq.(\ref{eq:Replica})] 
of the order of ${\cal O}[(1/N)^{1}]$ given by Fig.\ 7(b) in Ref.~\citen{Tsuruta5}. 
The explicit forms of higher order terms (dotted parts) 
are given by 
Figs.\ 8(c)-8(f) and Figs.\ 9(c)-9(m) in Ref.~\citen{Tsuruta5}. 
(c)  Self-energy from the lattice effect beyond the contribution from the local vertex
$\Gamma_{\rm loc}$ given by 
the {\it virtual} two-channel Anderson lattice model [Eq.(\ref{eq:Replica})].  
Its explicit form is essentially the same as that given in Ref.~\citen{Tsuruta6} 
except that the hybridization shown by dot ${\tilde V}=\sqrt{c_{\rm imp}}\,V$ 
has been modified in the 
{\it virtual} lattice system, and that the wave-vectors are defined on the {\it virtual} lattice.  
The structure of the renormalized vertex  $\Gamma_{\rm latt}$ is discussed in Appendix C. 
}
\label{Fig:SelfEnergy-1}
\end{center}
\end{figure}

The self-energies of conduction electrons in the {\it virtual} system consist of three parts, 
$\Sigma_{\rm c}^{(a)}$, $\Sigma_{\rm c}^{(b)}$, and $\Sigma_{\rm c}^{(c)}$, as shown in Fig.\ \ref{Fig:SelfEnergy-1}, 
in which $\Sigma_{\rm c}^{(a)}$ represents the effect of the two-channel Anderson impurity model, 
$\Sigma_{\rm c}^{(b)}$ the local effect from the lattice contribution given by the $d$-infinity lattice model, 
and $\Sigma_{\rm c}^{(c)}$ the contribution from the correction of the local vertex 
$\Gamma_{\rm loc}$, which is crucial in the finite dimensional 
two-channel Anderson lattice model as discussed in Ref.~\citen{Tsuruta6}.
The $T$ dependence of the imaginary part of 
these self-energies with the lowest correction in $T/{\tilde E}_0$ are given in Appendix C as follows: 
\begin{eqnarray}
& &
{\rm Im}\Sigma_{\rm c}^{(a){\rm R}}(\varepsilon=0;T)=
-{\tilde C}\left[1-2{\tilde a}_{\rm f}(T/{\tilde E}_{0})^{\nu}\left(1-\frac{1}{M^{2}}\right)\right]
+A_{\rm imp}T^{2}
\non\\
\label{ImSigma1}
\\
& &
{\rm Im}\Sigma_{\rm c}^{(b){\rm R}}(\varepsilon=0;T)={\tilde C}\frac{1}{M^{2}}
\label{ImSigma2}
\\
& &
{\rm Im}\Sigma_{\rm c}^{(c){\rm R}}(\varepsilon=0;T)={\tilde C}\,\frac{1-2{\tilde a}_{\rm f}
(T/{\tilde E}_{0})^{\nu}}{\displaystyle 1+\frac{{\tilde T}^{*}}{T}}
\frac{{\tilde T}^{*}}{T}
\left(1-\frac{1}{M^{2}}\right)
-A_{\rm latt}T^{2},
\non\\
\label{ImSigma3}
\end{eqnarray}
where $\nu\equiv (1-{\tilde a}_{\rm f})M/N$~\cite{Comment_nu}, 
and the explicit expressions of coefficients ${\tilde C}$, 
${\tilde a}_{\rm f}$, ${\tilde E}_{0}$, ${\tilde T}^{*}$, $A_{\rm imp}$, and $A_{\rm latt}$ 
are derived on the basis of discussion in relation to Eq.\ (40) in Ref.~\citen{Tsuruta6}
and that in Appendix C. 
However, it is crucial to note that the wave vector ${\bf p}$ is defined on the {\it virtual} lattice 
of the {\it virtual} system described by the Hamiltonian [Eq.\ (\ref{eq:Replica})] so that the hybridization 
is modified as ${\tilde V}\equiv \sqrt{c_{\rm imp}}V$ as discussed in Appendix B. 
Namely, the hybridization in the {\it virtual} 
system should be replaced by ${\tilde V}$ when we apply the results obtained in Ref.~\citen{Tsuruta6}.  
It is also crucial that ${\bf k}$-summation, $(1/N_{\rm L})\sum_{\bf k}$, in Ref.~\citen{Tsuruta6} 
should be replaced by $(1/N_{\rm f})\sum_{\bf p}$, i.e., 
\begin{eqnarray}
& &
\frac{1}{N_{\rm L}}\sum_{\bf k}F({\bf k})\Rightarrow\frac{1}{N_{\rm f}}\sum_{\bf p}F({\bf p})
=\frac{1}{c_{\rm imp}N_{\rm L}}\sum_{\bf p}F({\bf p})
\label{sum_virtual_lattice}
\end{eqnarray}
for an arbitrary function of $F({\bf p})$.

The coefficient  ${\tilde C}$ in Eqs.\ (\ref{ImSigma1})-(\ref{ImSigma3}) is given by Eq. (\ref{b8}) in Appendix C:
\begin{eqnarray}
& &
{\tilde C}=\frac{\pi}{N}\left(\frac{{\tilde a}_{\rm f}{\tilde V}^{2}}{{\tilde E}_{0}}\right)^{2}{\tilde N}(0), 
\label{eq:C}
\end{eqnarray}
where ${\tilde N}(0)$ is the spectral weight of conduction electrons defined by 
\begin{eqnarray}
& &
{\tilde N}(0)
\equiv-\frac{1}{\pi}\frac{1}{N_{\rm f}}\sum_{{\bf p},\tau}
{\rm Im}{\bar G}^{\rm R}_{\tilde {\rm c}\tau\sigma}({\bf p},0),
\label{eq:C1}
\end{eqnarray}
as derived from Eq.\ (40) in Ref.~\citen{Tsuruta6}. 
With the use of Eq.\ (\ref{sum_virtual_lattice}), ${\tilde N}(0)$ [Eq.\ (\ref{eq:C1})] is given as
\begin{eqnarray}
& &
{\tilde N}(0)=-\frac{1}{\pi}\frac{1}{c_{\rm imp}N_{\rm L}}
\sum_{{\bf p},\tau}{\rm Im}{\bar G}^{\rm R}_{{\tilde {\rm c}}\tau\sigma}({\bf p},0)
\equiv\frac{N(0)}{c_{\rm imp}}.
\label{eq:C2}
\end{eqnarray}
Therefore, ${\tilde C}$ [Eq.\ (\ref{eq:C})] is proportional to $c_{\rm imp}$ considering the relation 
${\tilde V}^{2}=c_{\rm imp}V^{2}$, because ${\tilde a}_{\rm f}$ and ${\tilde E}_{0}$ are independent of 
$c_{\rm imp}$ as discussed below. 

The residue ${\tilde a}_{\rm f}$ of the slave fermions in  Eqs.\ (\ref{ImSigma1}) and (\ref{ImSigma3})  
is given by Eq.\ (48) in Ref.~\citen{Tsuruta6} as 
\begin{eqnarray}
& &
\frac{1}{{\tilde a}_{\rm f}}=1+\frac{{\tilde V}^{2}}{N_{\rm f}}\sum_{{\bf p},\tau}\int d\varepsilon f(\varepsilon)\frac{-1}{\pi}
{\rm Im}{\bar G}_{\tilde {\rm c}\tau\sigma}({\bf p},\varepsilon)\frac{1}{(\varepsilon-{\tilde E}_{0})^{2}}, 
\label{eq:a_f}
\end{eqnarray}
where $f(\varepsilon)$ is the Fermi distribution function $f(\varepsilon)\equiv 1/(e^{\varepsilon/T}+1)$.

The binding energy ${\tilde E}_{0}$ of the slave fermions appearing 
in  Eqs.\ (\ref{ImSigma1}) and (\ref{ImSigma3}) is equivalent to the so-called Kondo temperature 
$T_{\rm K}$, and is given as a solution of the following equation derived in 
Ref.~\citen{Tsuruta6} [see Eq.\ (47)] 
\begin{eqnarray}
& &
\varepsilon_{\Gamma_{3}}-\varepsilon_{\Gamma_{7}}-{\tilde E}_{0}
-\frac{{\tilde V}^{2}}{N_{\rm f}}\sum_{{\bf p},\tau}\int d\varepsilon f(\varepsilon)
\frac{-1}{\pi}{\rm Im}{\bar G}^{\rm R}_{\tilde{\rm c}\tau\sigma}({\bf p},\varepsilon)
\frac{1}{\varepsilon-{\tilde E}_{0}}=0.
\label{eq:T_K}
\end{eqnarray}
In Eqs.\ (\ref{eq:a_f}) and (\ref{eq:T_K}), $\sigma$ dependence of ${\tilde a}_{\rm f}$ and 
${\tilde E}_{0}$ has been abbreviated because we are considering the para-magnetic state. 

The characteristic temperature ${\tilde T}^{*}$ in  Eq.\ (\ref{ImSigma3}) is given by
\begin{eqnarray}
& &
{\tilde T}^{*}=\frac{{\tilde E}_{0}}{1+{\tilde N}(0){\tilde E}_{0}}
\left[1-2{\tilde a}_{\rm f}(T/{\tilde E}_{0})^{\nu}\right],
\label{eq:T^*}
\end{eqnarray}
where the first factor in the limit $T=0$ was derived in Ref.~\citen{Tsuruta6} 
[see Eq.\ (59) and the following description], and the second factor in the bracket 
is derived in Appendix C [Eq.\ (\ref{b27})].  
In the low $T$ region, $T\ll {\tilde E}_{0}(0)$, ${\tilde T}^{*}$ is approximated by 
\begin{eqnarray}
& &
{\tilde T}^{*}\approx \frac{{\tilde E}_{0}(0)}{1+{\tilde N}(0){\tilde E}_{0}(0)},
\label{eq:T^*A}
\end{eqnarray}
which is the factor $bE_{0}$ in Eq.\ (59) of Ref.~\citen{Tsuruta6}.

With the use of ${\tilde C}$ [Eq.\ (\ref{eq:C})], 
as discussed in Appendix C [Eq.\ (\ref{b13})], the coefficient $A_{\rm imp}$ in Eq.\ (\ref{ImSigma1}) is given by
\begin{eqnarray}
& &
A_{\rm imp}\approx
\frac{\pi^{2}}{3}{\tilde C}[1-{\tilde a}_{\rm f}(0)][2+{\tilde a}_{\rm f}(0)]
\frac{1}{[{\tilde E}_{0}(0)]^{2}},
\label{A_imp}
\end{eqnarray}
while  
the coefficient $A_{\rm latt}$ in Eq.\ (\ref{ImSigma3}) is given, as discussed in Appendix C [Eq. (\ref{CImSigma3_A})] by 
\begin{eqnarray}
& &
A_{\rm latt}\approx\frac{\pi}{4}{\tilde C}^{2}N\frac{{\tilde a}_{\rm f}(0)}{[{\tilde E}_{0}(0)]^{3}}\frac{1}{M}, 
\label{A_latt}
\end{eqnarray}
where  ${\tilde a}_{\rm f}(0)$ and ${\tilde E}_{0}(0)$ are those values at $T=0$. 

Note here that ${\tilde C}$ is proportional to $c_{\rm imp}$ as discussed just below Eq.\ (\ref{eq:C2}), 
and that ${\tilde a}_{\rm f}$ and ${\tilde E}_{0}$ are independent of $c_{\rm imp}$ as discussed below. 
Therefore, the coefficient $A_{\rm imp}$ [Eq.\ (\ref{A_imp})] is proportional to 
$c_{\rm imp}$ reflecting the impurity effect, 
while the coefficient $A_{\rm latt}$ [Eq.\ (\ref{A_latt})] is proportional to
$[c_{\rm imp}]^{2}$.  
This $[c_{\rm imp}]^{2}$ dependence of $A_{\rm latt}$ can be understood from the structure of the 
Feynman diagram shown in Fig.\ \ref{Fig:SelfEnergy-1}(c). 
Namely, two internal wave-vectors summations give a factor 
$[c_{\rm imp}]^{-2}$ according to the rule of Eq.\ (\ref{sum_virtual_lattice}) while eight hybridizations 
${\tilde V}^{8}$ give a factor $[c_{\rm imp}]^{4}$ because of the relation 
${\tilde V}=\sqrt{c_{\rm imp}}\,V$ in the {\it virtual} system. 
This result is also plausible because the Feynman diagram of Fig.\ \ref{Fig:SelfEnergy-1}(c) represents 
the effect of inter-site of Pr ions in the original lattice shown in Fig.\ \ref{Fig:Replica}(a) and is 
considered to be proportional to the product of probability $c_{\rm imp}$ of Pr ions in the original 
lattice system.  

Adding the contributions [Eqs.\ (\ref{ImSigma1})-(\ref{ImSigma3})] together with 
the weak damping effect by static scattering of conduction electrons by localized f electrons at Pr sites, 
the imaginary part of the retarded self-energy of the conduction electrons 
Im$\Sigma^{\rm R}_{c}(\varepsilon=0;T)$ is given 
as
\begin{eqnarray}
& &
{\rm Im}\Sigma_{\rm c}^{\rm R}(\varepsilon=0;T)
=
-{\tilde C}\left[1-2{\tilde a}_{\rm f}(T/{\tilde E}_{0})^{\nu}\right]\frac{T}{\displaystyle T+{\tilde T}^{*}}
\left(1-\frac{1}{M^{2}}\right)
\nonumber
\\
& &
\qquad\qquad\qquad\qquad
+A_{\rm imp}T^{2}-A_{\rm latt}T^{2}-\frac{1}{2{\tilde \tau}_{\rm imp}},
\label{ImSigmaTotal}
\end{eqnarray}
where $1/2{\tilde \tau}_{\rm imp}$ is the impurity scattering rate of the conduction electrons 
renormalized by the lattice effect in general, although it is essentially unrenormalized
as discussed in Sect. 5.

The $c_{\rm imp}$ dependence of the 
above coefficients is important for the discussions below, and is given as follows.  
First of all,  ${\tilde a}_{\rm f}$ and ${\tilde E}_{0}$ are independent of $c_{\rm imp}$. 
This is because equations determining these quantities, Eqs.\ (\ref{eq:a_f}) and (\ref{eq:T_K}), 
are transformed to Eqs.\ (48) and (47) in Ref.~\citen{Tsuruta6}, respectively, 
considering that ${\tilde V}=\sqrt{c_{\rm imp}}\,V$ and the relation Eq.\ (\ref{sum_virtual_lattice}) 
holds.  
Since ${\tilde a}_{\rm f}$ and ${\tilde E}_{0}$ are independent of $c_{\rm imp}$, 
the coefficient $A_{\rm imp}$ [Eq.\ (\ref{A_imp})] is proportional to $c_{\rm imp}$ while 
the coefficient $A_{\rm latt}$ [Eq.\ (\ref{A_latt})] is proportional to $[c_{\rm imp}]^{2}$, as mentioned 
above. 
On the other hand, ${\tilde T}^{*}$ given by Eq.\ (\ref{eq:T^*A}) has no simple power-law dependence 
on $c_{\rm imp}$. Indeed, using the expression for ${\tilde N}(0)$ [Eq.\ (\ref{eq:C2})], 
${\tilde T}^{*}$ [Eq.\ (\ref{eq:T^*A})] is reduced to 
\begin{eqnarray}
& &
{\tilde T}^{*}\approx \frac{c_{\rm imp}{\tilde E}_{0}(0)}{c_{\rm imp}+N(0){\tilde E}_{0}(0)}.
\label{eq:T^*B}
\end{eqnarray}
Namely, ${\tilde T}^{*}\approx c_{\rm imp}/N(0)\sim c_{\rm imp}/N_{\rm F}$ in the low concentration limit 
of Pr ions, i.e., $c_{\rm imp}\lsim N(0){\tilde E}_{0}(0)$, while  ${\tilde T}^{*}\sim {\tilde E}_{0}(0)$ 
in a wide rage of concentration 
$c_{\rm imp}\gsim N(0){\tilde E}_{0}(0)\sim N_{\rm F}{\tilde E}_{0}(0)$ which is far smaller than 1 
because ${\tilde E}_{0}(0)\sim T_{\rm K}$ is far smaller than the Fermi energy $E_{\rm F}$ of conduction 
electrons in the present situation. 
\section{Temperature Dependence of Physical Quantities}
In this section, we discuss the $T$ dependence of various physical quantities. 

With the use of Im$\Sigma_{\rm c}^{\rm R}(\varepsilon=0;T)$ [Eq.\ (\ref{ImSigmaTotal})], 
the $T$ dependence of the resistivity is essentially given by 
\begin{eqnarray}
\rho(T)=-\frac{2m}{N_{\rm e}e^{2}}
{\rm Im}\Sigma_{\rm c}^{\rm R}(\varepsilon=0;T),
\label{Rho_Temp}
\end{eqnarray}
where 
$m$ and $N_{\rm e}$ are the mass of the free electron and the number density of conduction electrons, 
respectively, and 
we have assumed that the dispersion of conduction electrons is given by 
that of the free electron.  Therefore, the $T$ dependent part is essentially the same as that of 
the bulk pure system 
except for the $c_{\rm imp}$ dependence arising from the factor ${\tilde C}\propto c_{\rm imp}$ 
and ${\tilde T}^{*}$ [Eq.\ (\ref{eq:T^*B})]~\cite{Comment0}.   
This is because the effect of the bare impurity scattering rate $1/2\tau_{\rm imp}$ in 
the Green function of conduction electrons [Eq.\ (\ref{Green_c_p:2})] does not alter 
the fundamental structure of $T$ dependence in 
Im$\Sigma^{{\rm R}}_{\rm c}(\varepsilon=0;T)$ 
[Eq.\ (\ref{ImSigmaTotal})].
Therefore, the so-called scaling behavior of the $T$ dependence in $[\rho(T)-\rho_{0}^{*}]$, 
i.e., that given by Eq.\ (\ref{Rho_lattice}), is expected to hold in the rather wide temperature region 
$T<{\tilde T}^{*}$ as in bulk pure systems~\cite{Tsuruta6}.  

One might think that the residual part at $T=0\,$K given by 
${\rm Im}\Sigma^{{\rm R}}_{\rm c}(\varepsilon=0;0)$, the renormalized scattering rate 
$1/2{\tilde \tau}_{\rm imp}$ in Eq.\ (\ref{ImSigmaTotal}), gives some additional 
residual resistivity other than that from $1/2\tau_{\rm imp}$. 
However, we have verified that ${\tilde \tau}_{\rm imp}$ is not influenced by a direct numerical 
calculation of Im$\Sigma^{{\rm R}}_{\rm c}(\varepsilon=0;T=0)$. 
In the case of the Anderson model with  the zero interaction between f-electrons, $U_{ff}=0$, $1/2{\tilde \tau}_{\rm imp}\propto c_{\rm imp}(1-c_{\rm imp})$ because of Nordheim rule.
Even if we increase $U_{ff}$ adiabatically, the $c_{\rm imp}$ dependence of $1/2{\tilde \tau}_{\rm imp}$ does not change.
In the numerical calculation in \S6, 

The chemical potential $\mu(T)$ and the specific heat $C_{V}(T)$ are also expected to 
exhibit the same $T$ dependence as the bulk pure system because they are essentially determined by the 
{\it virtual} Hamiltonian [Eq.\ (\ref{eq:Replica})].
Namely, they are expected to exhibit the following  $T$ dependence 
in the region 
$T_{\rm Q}<T\lsim {\tilde E}_{0}$, with $T_{\rm Q}$ being the transition temperature of the quadrupole
ordering~\cite{Tsuruta6}:
\begin{eqnarray}
\mu(T)\propto {\rm const.}-\sqrt{T},  
\label{mu(T)}
\end{eqnarray}
and 
\begin{eqnarray}
C_{V}(T)\propto {\rm const.}-\sqrt{T}.
\label{C_V(T)}
\end{eqnarray}
Such a $T$ dependence in the specific heat does not contradict the observation 
in Y$_{1-x}$Pr$_x$Ir$_2$Zn$_{20}$ ($x=0.024,\,0.044,\,0.085,$ and $ \,0.44$).
{
This is because the data on the $T$ dependence reported in Ref.~\citen{Yamane} can be fitted also by the functional form of Eq. (\ref{C_V(T)}) in the finite $T$ range, although they fitted them by 
$C(T)\propto  -T\log\,T$~\cite{Comment}.
}

\section{Difference from Single-Channel Kondo Impurities System} 
\begin{figure}[h]
\begin{center}
\includegraphics[width=0.6\linewidth]{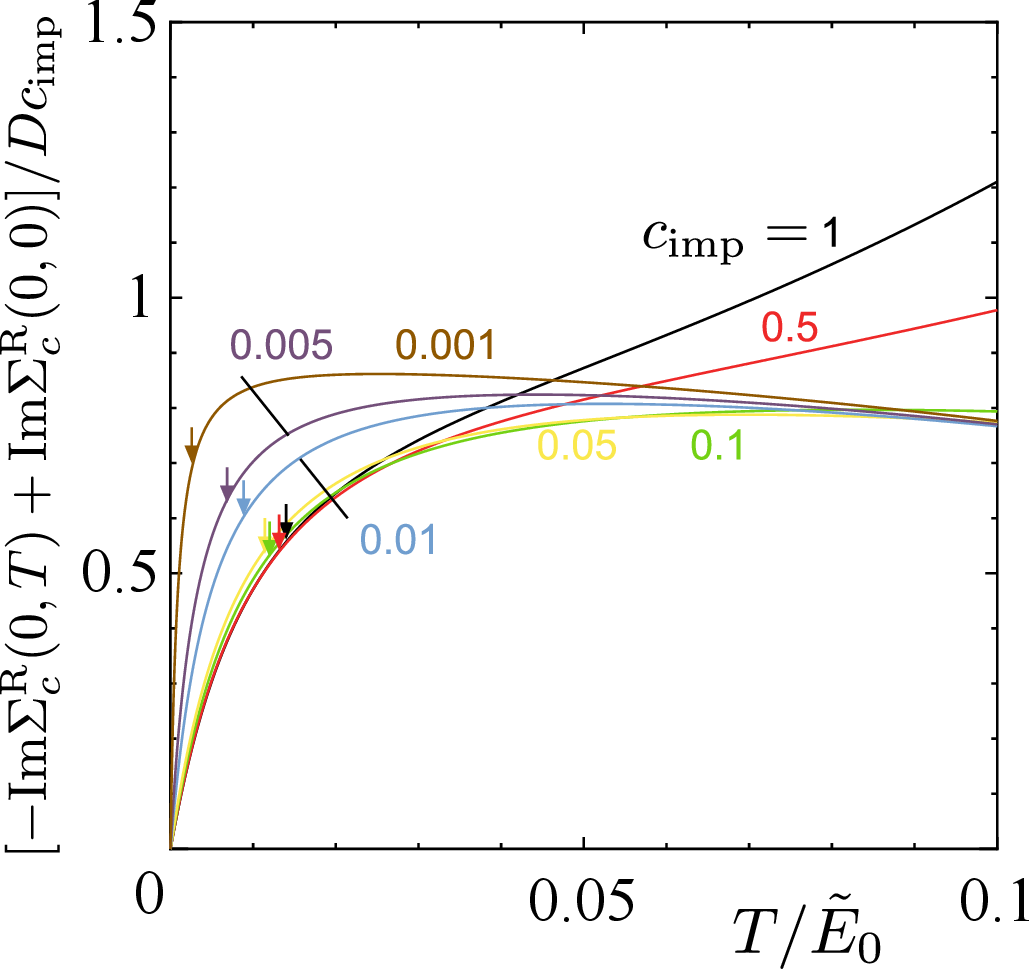}
\caption{
$[-{\rm Im}\Sigma_{\rm c}^{\rm R}(0,T)+
{\rm Im}\Sigma_{\rm c}^{\rm R}(0,0)]$ in Eq.\ (\ref{ImSigmaTotal}) as a unit of 
$Dc_{\rm imp}$ for $M=2$ and the parameter set, 
$\tilde{V}/D=0.3, (\varepsilon_{\Gamma_3}-\varepsilon_{\Gamma_7})/D=-0.4, \tilde{E}_0/D=0.0117$, and $\tilde{a}_f=0.115$,
as a function of  $T/{\tilde E}_{0}$ for a series of $c_{\rm imp}=1\sim 0.001$.  
Arrows indicate the temperature below which $\sqrt{T}$-like behavior appears down to $T= {\tilde T}^{*}$.
}
\label{Fig:Resistivity-1}
\end{center}
\end{figure}

As discussed in the previous sections, the $T$ dependence of the resistivity in the 
two-channel Anderson impurities model is clearly different from that of the single-impurity two-channel 
Kondo~\cite{Cox3,Cox2} or Anderson model~\cite{Tsuruta2} in which 
the resistivity is proportional to (const.$-\sqrt{T}$) toward $T=0$ 
in the weak-coupling case{~\cite{Ludvig}}.  
On the other hand, in the present case, the resistivity 
arising from Im$\Sigma_{\rm c}^{\rm R}(\varepsilon=0;T)$ [Eq.\ (\ref{ImSigmaTotal})] 
does not increase toward $T=0$ even though there exists a factor 
$[1-2{\tilde a}_{\rm f}(T/{\tilde E}_{0})^{\nu}]$ that gives the non-Fermi liquid behavior expected in 
the single-impurity two-channel Kondo effect~\cite{Tsuruta2,Comment_nu}. 
Indeed, the quantity $[-{\rm Im}\Sigma_{\rm c}^{\rm R}(0,T)+
{\rm Im}\Sigma_{\rm c}^{\rm R}(0,0)]/Dc_{\rm imp}$ in Eq.\ (\ref{ImSigmaTotal}), which is proportional to the resistivity, 
for the parameter set, 
$\tilde{V}/D=0.3, (\varepsilon_{\Gamma_3}-\varepsilon_{\Gamma_7})/D=-0.4, \tilde{E}_0/D=0.0117,$ and $\tilde{a}_f=0.115$, 
is shown in Fig.\ \ref{Fig:Resistivity-1} for a series of $c_{\rm imp}$.
One can see that the scaling behavior 
in $T$ dependence of  $[\rho(T)-\rho_{0}^{*}]/c_{\rm imp}$ [Eq.\ (\ref{Rho_lattice})] holds 
down to the low concentration $c_{\rm imp}\simeq 0.001$ less than that attained experimentally 
so far~\cite{Yamane}.

This is because the presence of a 
factor $T/(T+{\tilde T}^{*})$ in Eq.\ (\ref{ImSigmaTotal}) 
invalidates the increase in the resistivity, 
given by the factor $[1-2{\tilde a}_{\rm f}(T/{\tilde E}_{0})^{\nu}]$,
in the low $T$ region $T<{\tilde T}^{*}$. 
Therefore, in order that the single-impurity behavior is observed, ${\tilde T}^{*}$ should be extremely 
small, e.g., ${\tilde T}^{*}=0.01\,$K which is a typical lower limit of $T$ in the standard low 
temperature measurements using the dilution refrigerator.  
By solving approximate relation [Eq. (\ref{eq:T^*B})], 
the concentration $c_{\rm imp}$ is expressed by a function of 
${\tilde T}^{*}$ as 
\begin{eqnarray}
	& &
	c_{\rm imp}\approx
	\frac{{ \tilde T}^{*}N(0){\tilde E}_{0}(0)}
	{ {\tilde E}_{0}(0)-{\tilde T}^{*}}.
	\label{eq:c_imp}
\end{eqnarray}
Since ${\tilde E}_{0}(0)$ or the Kondo temperature $T_{\rm K}$ is of the order of 10K in a conventional heavy fermion system,
and $N(0) \sim 1/D$ with $D$ being half the bandwidth of conduction electrons
of the order of $10^{4}$K in a typical metal, 
the critical concentration $c_{\rm imp}^{\rm cr}$, below which the 
$[1-2{\tilde a}_{\rm f}(T/{\tilde E}_{0})^{\nu}]$ like $T$ dependence in the resistivity 
is expected to be observed around $T\gsim {\tilde T}^{*}\sim 10^{-2}\,$K, is roughly estimated as 
\begin{eqnarray}
c_{\rm imp}^{\rm cr}\sim \frac{{\tilde T}^{*}}{D}\sim 10^{-6},
\label{c^{cr}_{imp}}
\end{eqnarray}
where we have neglected ${\tilde T}^{*}$  compared to ${\tilde E}_{0}(0)$ in the 
denominator of (Eq.\ (\ref{eq:c_imp}) because we are interested in the case 
where ${\tilde T}^{*}\sim 10^{-2}\,$K. 
Thus, it is extremely difficult to observe experimentally the non-Fermi liquid $T$ 
dependence of the resistivity predicted on the 
single-impurity two-channel Kondo effect~\cite{Cox3,Cox2}.  

The physical reason for this extremely small $c_{\rm imp}^{\rm cr}$ may be traced back to 
the character of the two-channel Kondo effect in which the local moment cannot be 
effectively screened out with the finite range of ${\cal O}(aD/T_{\rm K})$ 
with $a$ being the mean distance among conduction electrons, but is over-screened 
unlike in the case of the single-channel Kondo effect~\cite{Nozieres,Comment_xi}, 
resulting in the screening length (if any) diverges or extremely long 
compared to the case of the single-channel Kondo effect.

This situation is in marked contrast to the case of the single-channel ($M=1$) Kondo or Anderson 
impurities model in which the $T$ dependence of the resistivity at $T\ll T_{\rm K}$ 
follows (const.$-T^{2}$) dependence up to the relatively large concentration 
$c_{\rm imp}^{\rm cr}\sim 0.5$ of f-ions, e.g., Ce, as observed in Ce$_{x}$La$_{1-x}$Cu$_6$~\cite{Sumiyama,Onuki}.     
The difference stems from that of the $M^{2}$ dependence in the expression [Eq.\ (\ref{ImSigmaTotal})]. 
Namely, the anomalous $T$ dependence with $M\ge 2$ disappears in the case of single-channel with $M=1$. 
Therefore, Eq.\ (\ref{ImSigmaTotal}) is reduced to 
\begin{eqnarray}
& &
{\rm Im}\Sigma_{\rm c}^{\rm R}(\varepsilon=0;T)=
(A_{\rm imp}-A_{\rm latt})T^{2}-\frac{1}{2{\tilde \tau}_{\rm imp}}.
\label{ImSigmaTotal_Kondo}
\end{eqnarray}
It is crucial to note that $A_{\rm imp}$ is proportional to $c_{\rm imp}$ while $A_{\rm latt}$ 
is proportional to $[c_{\rm imp}]^{2}$ as Eqs.\ (\ref{A_imp}) and (\ref{A_latt}), respectively, 
because ${\tilde C}$ [Eq.\ (\ref{eq:C})] is proportional to $c_{\rm imp}$.
This implies that, in the low concentration region ($c_{\rm imp}\ll1$), 
the sign of $T^{2}$ term in Im$\Sigma_{\rm c}^{\rm R}(\varepsilon=0;T)$ is positive or 
the sign of $T^{2}$ term in $\rho(T)$ is negative, 
which leads to the local-Fermi liquid behavior~\cite{Nozieres2}. 
On the other hand, in the high concentration region ($c_{\rm imp}\lsim 1$), the heavy Fermi liquid 
behavior is realized. This aspect is consistent with the observation reported in Ref.~\citen{Onuki}.  

\begin{figure}[h]
\begin{center}
\rotatebox{0}{\includegraphics[width=0.6\linewidth]{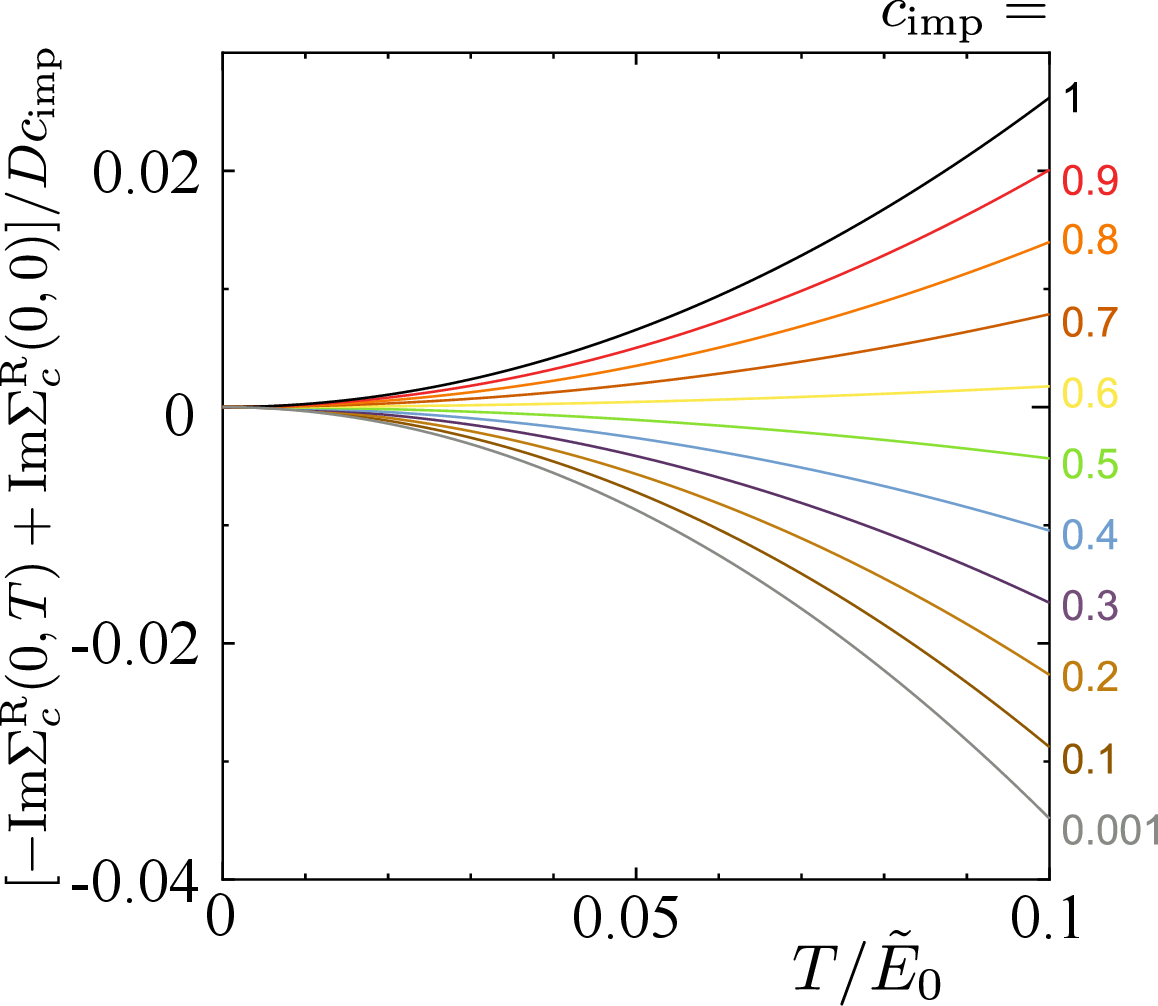}}
\caption{
$[-{\rm Im}\Sigma_{\rm c}^{\rm R}(0,T)+{\rm Im}\Sigma_{\rm c}^{\rm R}(0,0)]$ in Eq.\ (\ref{ImSigmaTotal}) 
as a unit of $Dc_{\rm imp}$ for $M=1$ with the parameter set, 
$\tilde{V}/D=0.4$, $\varepsilon_{\Gamma_3}/D=0$, $\varepsilon_{\Gamma_7}/D=-0.4$, $\tilde{E}_0/D=0.0821$, and $\tilde{a}_{\rm f}=0.339$,
as a function of  $T/{\tilde E}_{0}$ for a series of $c_{\rm imp}=1\sim 0.001$. 
The coefficient of $T^{2}$ dependence changes sign from positive to negative between $c_{\rm imp}=0.6$ and $c_{\rm imp}=0.5$ 
as decreasing the concentration 
$c_{\rm imp}$, which is consistent with 
the observation reported in Ce$_x$La$_{1-x}$Cu$_6$~\cite{Sumiyama,Onuki}.}
\label{Fig:Resistivity-2}
\end{center}
\end{figure}

The critical concentration $c_{\rm imp}^{\rm cr}$, which separates the two Fermi liquid behaviors,  
is given by the condition $A_{\rm imp}=A_{\rm latt}$. Namely, by equating the expressions 
$A_{\rm imp}$ [Eq.\ (\ref{A_imp})] and $A_{\rm latt}$ [Eq.\ (\ref{A_latt})], with the use of 
Eqs.\ (\ref{eq:C}) and (\ref{eq:C2}) and the relation ${\tilde V}=\sqrt{c_{\rm imp}}\,V$, it is given as 
\begin{eqnarray}
& &
c_{\rm imp}^{\rm cr}\approx
\frac{8}{3}\left(\frac{{\tilde E}_{0}(0)}{{\tilde a_{\rm f}}(0)V^{2}}\right)^{2}
\frac{1}{N(0)}\frac{{\tilde E}_{0}(0)}{{\tilde a}_{\rm f}(0)},
\label{c_{imp}^{cr}}
\end{eqnarray}
where we have approximated as $[1-{\tilde a}_{\rm f}(0)][2+{\tilde a}_{\rm f}(0)]\approx 2$ 
in the expression of $A_{\rm imp}$ [Eq.\ (\ref{A_imp})] because 
${\tilde a}_{\rm f}(0)\ll 1$. 
Since ${\tilde E}_{0}(0)\sim {\tilde a}_{\rm f}(0)V^{2}/D$ so that 
${\tilde E}_{0}(0)/{\tilde a}_{\rm f}(0)\sim V^{2}/D$ according to the periodic Anderson model 
picture~\cite{Rice-Ueda},  the critical concentration $c_{\rm imp}^{\rm cr}$ is roughly estimated as  
\begin{eqnarray}
& &
c_{\rm imp}^{\rm cr}\sim 
\frac{8}{3}\left(\frac{V}{D}\right)^{2}\,\frac{1}{N(0)D}\sim\frac{8}{3}\left(\frac{V}{D}\right)^{2}. 
\label{c_{imp}^{cr}2}
\end{eqnarray}
This is not extremely smaller than 1 in the usual situation for heavy Fermion metals.
The result for the quantity $[-{\rm Im}\Sigma_{\rm c}^{\rm R}(0,T)+
{\rm Im}\Sigma_{\rm c}^{\rm R}(0,0)]/Dc_{\rm imp}$ in Eq.\ (\ref{ImSigmaTotal_Kondo}) 
with the parameter set,
$\tilde{V}/D=0.4$, $\varepsilon_{\Gamma_3}/D=0$, $\varepsilon_{\Gamma_7}/D=-0.4$, $\tilde{E}_0/D=0.0821$, and $\tilde{a}_{\rm f}=0.339$,
is shown in Fig.\ \ref{Fig:Resistivity-2} 
for a series of $c_{\rm imp}$, in which one can see that the coefficient of 
the $T^{2}$ term changes the sign at between $c_{\rm imp}=0.5$ and 
$c_{\rm imp}=0.6$. This is consistent with the experiment of the $T$ dependence in the 
resistivity of Ce$_x$La$_{1-x}$Cu$_6$~\cite{Sumiyama,Onuki}. 

{
Furthermore, the critical concentration $c_{\rm imp}^{\rm cr}\sim 0.5$ in the single-channel impurities Anderson model is also consistent with the theoretical result of the $T$ dependence in the resistivity
of Ce$_x$La$_{1-x}$Cu$_6$ obtained with the use of the dynamical mean field theory (DMFT) and the coherent potential approximation (CPA)~\cite{Mutou}.
However, the result of Ref.~\citen{Mutou} deviates from the Nordheim rule, $\rho\propto x(1-x)$, in the region $x<0.2$ probably due to the difficulty of CPA when applied to the impurities Anderson model as mentioned in the last paragraph in $\S$3.
In addition, the $T$ dependence in the resistivity in the two-channel lattice system
and the two-channel impurities system cannot be explained by DMFT
because the inter-site effect, which is crucial in the case of the two-channel Anderson models, lattice or impurities, is not taken into account in the formalism of DMFT.
}

Concluding this section, let us remark on the relation between the expression for 
$A_{\rm latt}$ [Eq.\ (\ref{A_latt})] and Eq.\ (57) in Ref.~\citen{Tsuruta6}.  
According to the formula [Eq.\ (\ref{Rho_Temp})], the Fermi liquid component of the resistivity 
$\rho_{\rm FL}$ is given by 
\begin{eqnarray}
\rho_{\rm FL}\approx\frac{2m}{N_{\rm e}e^2}A_{\rm latt}T^{2}.
\label{Rho_FL}
\end{eqnarray} 
Substituting the expression for $A_{\rm latt}$ [Eq.\ (\ref{A_latt})] with the expression 
${\tilde C}$ [Eq.\ (\ref{eq:C})], Eq.\ (\ref{Rho_FL}) is reduced to 
\begin{eqnarray}
\rho_{\rm FL}=\frac{2m}{N_{\rm e}e^2}
\frac{\pi^{3}}{4N}\left(\frac{{\tilde a}_{\rm f}V^{2}}{{\tilde E}_{0}}\right)^{4}
[N(0)]^{2}\frac{{\tilde a}_{\rm f}}{{\tilde E}_{0}^{3}}c_{\rm imp}^{2}T^{2},  
\label{Rho_FL2}
\end{eqnarray} 
where ${\tilde a}_{\rm f}(0)$ and ${\tilde E}_{0}(0)$ have been abbreviated by ${\tilde a}_{\rm f}$ 
and ${\tilde E}_{0}$. With the use of approximate relations, ${\tilde E}_{0}\sim {\tilde a}_{\rm f}V^{2}/D$,  
$N(0)D\sim 1$, and $N(0)\sim N_{\rm F}$, Eq.\ (\ref{Rho_FL2}) is reduced to 
\begin{eqnarray}
\rho_{\rm FL}=\left[\frac{2m}{N_{\rm e}e^2}
\frac{\pi^{3}}{4N}\left(\frac{D}{V}\right)^{4}c_{\rm imp}^{2}\right]\,
\pi N_{\rm F}V^{2}\frac{T^{2}}{{\tilde E}_{0}^{2}}. 
\label{Rho_FL3}
\end{eqnarray} 
Comparing this expression with that of Eq.\ (57) in Ref.~\citen{Tsuruta6}, the factor in the 
brackets of Eq.\ (\ref{Rho_FL3}) should be identified by a factor $r$ in Eq.\ (57) of Ref.~\citen{Tsuruta6} 
for the bulk pure system, i.e., $c_{\rm imp}=1$.  


\section{Conclusion}
We have shown theoretically that the two-channel Anderson impurities system as 
Y$_{1-x}$Pr$_x$Ir$_2$Zn$_{20}$ ($x=0.024,\,0.044,\,0.085,\,$ and $0.44$) exhibits essentially 
the same non-Fermi liquid behaviors as 
the pure system with $x=1$ unless $x$ is extremely small less than $x^{\rm cr}\sim 10^{-6}$ 
for a reasonable set of parameters. It was crucial to introduce a new formalism of treating the effect 
of random distribution of dilute Pr ions on virtual periodic lattice system. 
On this formalism, the theory for the lattice system can be applied with modifications of 
relevant parameters in the pure lattice system almost as it stands. In particular, 
the $T$ dependence 
of the resistivity so calculated explains quite well that observed experimentally in diluted system 
Y$_{1-x}$Pr$_x$Ir$_2$Zn$_{20}$.  
The $T$ dependence of other physical quantities, such 
as the specific heat, are also the same as those in periodic lattice systems.
The critical impurities concentration $c_{\rm imp}^{\rm cr}$, below which the resistivity 
shows the temperature dependence of the single two-channel impurity model, has been shown 
to be extremely small not reached by controlled experiments, while that for the single-channel model 
is only moderately smaller than 1 in consistent with the observation in Ce-based impurity heavy 
fermion systems such as Ce$_x$La$_{1-x}$Cu$_6$.

\section*{Acknowledgments}
We benefited much from stimulating conversations with K. Izawa, Y. Yamane and T. Onimaru for 
experimental aspects of the diluted Pr-1-2-20 system.  
This work is supported by the Grant-in-Aid for Scientific Research 
(No. 17K05555) from the Japan Society for the Promotion of Science. 

\newpage
\appendix
\section{Comment on Scaling Behavior of Temperature Dependence of Resistivity}
In this appendix, we discuss the relationship between the $T$ 
dependence of the resistivity $\rho(T)$ [Eq.\ (\ref{rho_T})] and its scaling $T$ 
dependence given by analyses 
of experiments for Pr-1-2-20 systems performed in Refs.~\citen{Onimaru3,Yoshida_Izawa}. 

First,  we should note that the definition of $T_{0}$ in Eq.\ (\ref{rho_T}) of the present paper 
is different from that of $T^{(\rho)}_{0}$ in Ref.~\citen{Onimaru3} (and $T_{0}^{(17)}$ in Ref.~\citen{Yoshida_Izawa}), in which $T^{(\rho)}_{0}$ (and $T_{0}^{(17)}$) are determined by the condition that 
the $T$ dependence of the resistivity is scaled by $T_{0}^{(\rho)}$ (and $T_{0}^{(17)}$) with choosing the 
coefficient $b$ ($a_2$ in their notation) appropriately. Here we have redefined 
``$T_{0}$'' in Ref.~\citen{Yoshida_Izawa} as $T_{0}^{(17)}$ 
to distinguish it from $T_{0}$ defined in the present paper as the temperature where $T$ dependence in 
$\rho(T)$ starts to apparently deviate from $\sqrt{T}$ dependence.  
According to Fig.\ 3(a) in Ref.~\citen{Yoshida_Izawa}, 
$T_{0}^{(17)}\simeq T_{0}/0.45$ with $a_{2}=0.3$ for PrRh$_2$Zn$_{20}$ 
and 
$T_{0}^{(\rho)}\simeq T_{0}/0.75$ with $a_{2}=0.5$ for PrIr$_2$Zn$_{20}$, 
both of which correspond to taking $b\simeq 0.67$ and 0.67, respectively, 
in Eq.\ (\ref{rho_T}). Note that the coefficient $b$'s for both cases are the same within 
experimental errors, supporting that the scaling behavior is universal.    

\begin{figure}[h]
\begin{center}
\rotatebox{0}{\includegraphics[width=0.6\linewidth]{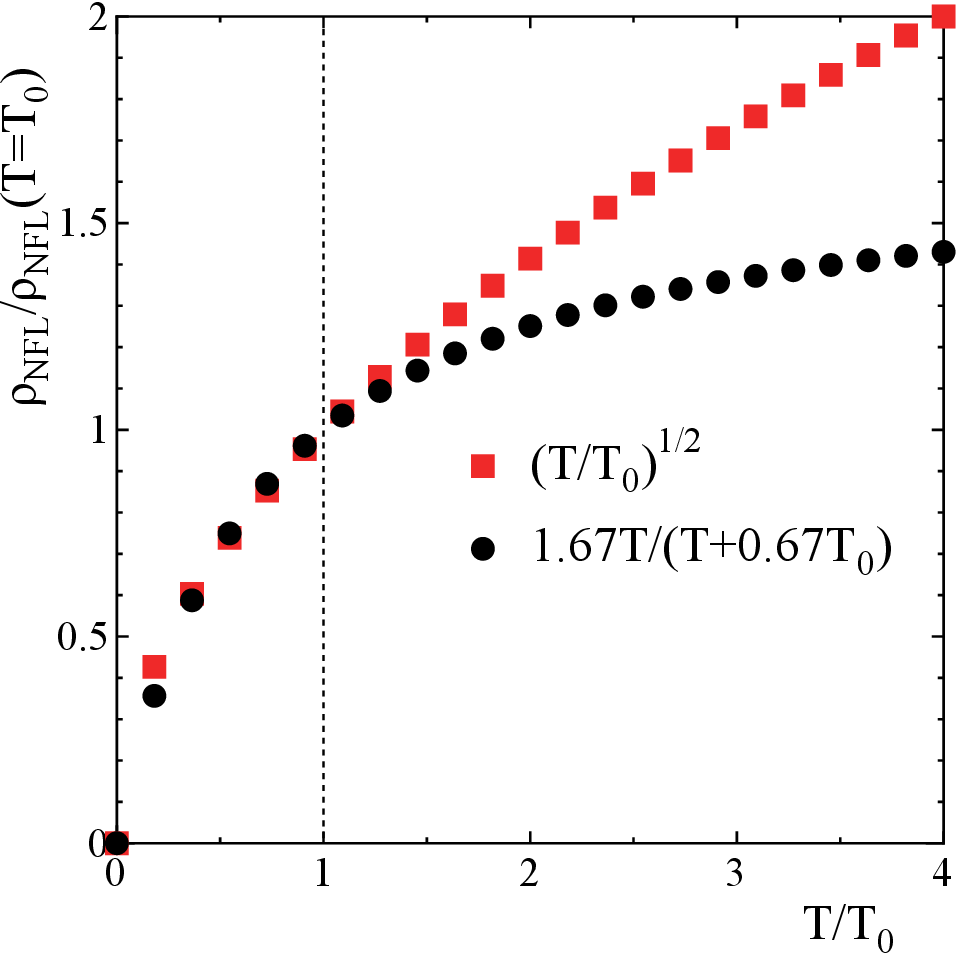}}
\caption{
Black circles and red squares represent data points of the expression $T/(T+bT_{0})$, 
the coefficient of the first term in Eq.\ (\ref{rho_T}) with $a=1$ and $b=0.67$, 
and those of the expression $\sqrt{T/T_{0}}/(1+b)$, respectively. The abscissa is $T/T_{0}$.  
}
\label{Fig:Rho_Theory}
\end{center}
\end{figure}

{Furthermore,} the coefficient $b\simeq 0.67$ is also consistent with the theoretical expression 
[Eq.\ (\ref{rho_T})] as seen in Fig.\ \ref{Fig:Rho_Theory} in which two expressions, 
$T/(T+bT_{0})$ and $\sqrt{T/T_{0}}/(1+b)$, are drawn as a function of $T/T_{0}$. 
Indeed, one can see that the theoretical expression $T/(T+bT_{0})$ with $b=0.67$ starts to 
apparently deviate from $\sqrt{T}$ dependence at $T/T_{0}\simeq 1.00$, 
which well mimics the behaviors observed in Refs.~\citen{Onimaru3,Yoshida_Izawa} (see Fig\ 3(a) in Ref.~\citen{Yoshida_Izawa}). 

\begin{figure}[h]
\begin{center}
\rotatebox{0}{\includegraphics[width=0.6\linewidth]{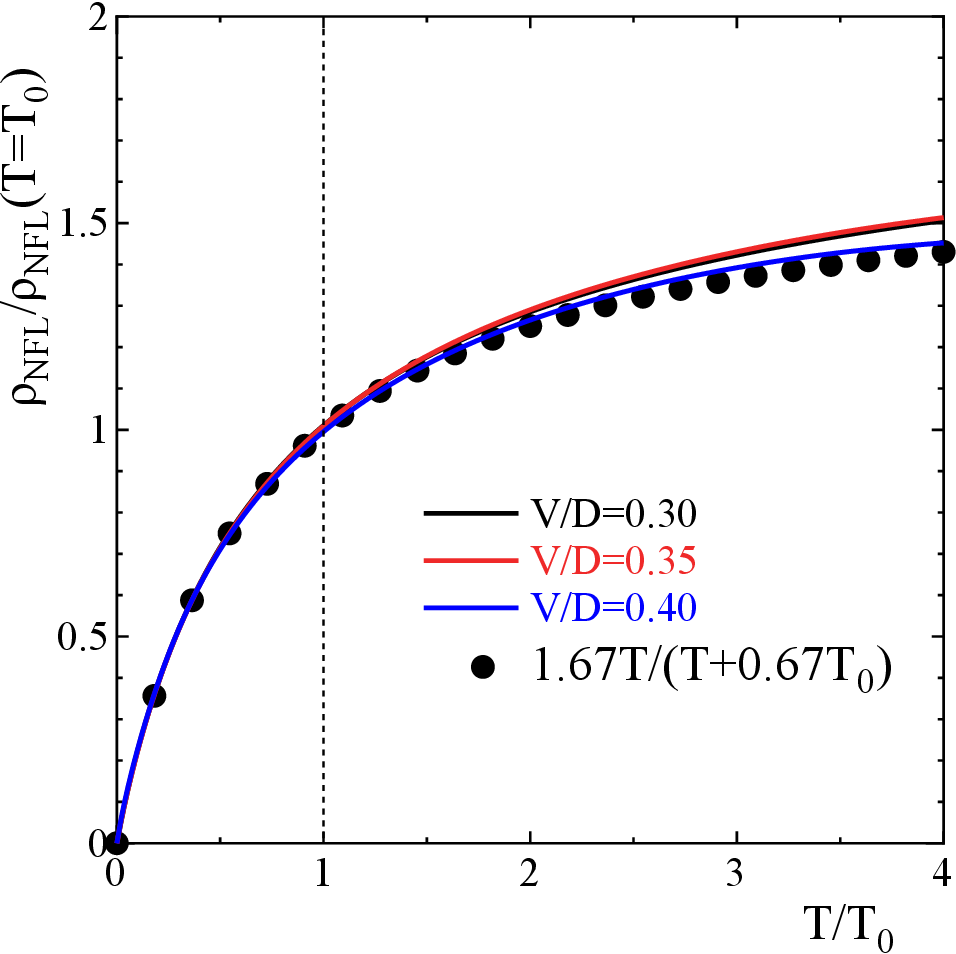}}
\caption{
Relationship between {\it approximate} expression [Eq.\ (\ref{rho_T})] 
with $b=0.67$ (black circles) and 
numerical result for $\rho_{\rm NFL}(T)$ given by Eq.\ (58) and 
presented in Fig.\ 11(c) of Ref.~\citen{Tsuruta6} for a series of 
c-f hybridizations $V/D=0.30$, $0.35$, and $0.40$. 
Note that the abscissa is $T/T_{0}$ and is different from 
$T/T_{0}^{\rm TM}$ in Fig.\ 11(c) of 
Ref.~\citen{Tsuruta6} as discussed in the text. 
}
\label{Fig:Tsuruta_Miyake}
\end{center}
\end{figure}

Concluding this appendix, we remark the relationship between the {\it approximate} expression 
[Eq.\ (\ref{rho_T})], which is essentially the same as the second term in Eq.\ (59) in 
Ref.~\citen{Tsuruta6}, and the numerical results for $\rho_{\rm NFL}(T)$ given by 
Eq.\ (58) and shown in Fig.\ 11(b) of Ref.~\citen{Tsuruta6}. It is crucial to note that the 
characteristic temperature $T_{0}$ (denoted by $T_{0}^{\rm TM}$ hereafter) in Ref.~\citen{Tsuruta6} 
is defined as the temperature where $\rho_{\rm NFL}(T)$ starts to apparently deviate 
from $({\rm const.}+\sqrt{T/D})$ behavior in Fig.\ 11(c), 
while $T_{0}$ of the present paper is defined as the temperature where it starts to apparently 
deviate from $\sqrt{T}$-like behavior without the ``const.'', i.e., the offset 
introduced for extracting the component proportional to $\sqrt{T}$ in Ref.~\citen{Tsuruta6}.   
In Fig.\ \ref{Fig:Tsuruta_Miyake}, the expression [Eq.\ (\ref{rho_T})] with $b=0.67$, and 
numerical result for $\rho_{\rm NFL}(T)$ given by Eq.\ (58) and presented in Fig.\ 11(c) of 
Ref.~\citen{Tsuruta6}, are shown as functions of $T/T_{0}$ instead of $T/T_{0}^{\rm TM}$. 
One can see that both curves almost coincide with each other, implying that the scaling behaviors 
confirmed in PrRh$_2$Zn$_{20}$~\cite{Yoshida_Izawa} and PrIr$_2$Zn$_{20}$~\cite{Onimaru3} 
agree with the prediction given by Ref.~\citen{Tsuruta6}.

\section{c-f Hybridization in {\it virtual} lattice system}
In this appendix, we derive the hybridization term ($\equiv H^{\rm v}_{\rm hyb}$) 
in the {\it virtual} Hamiltonian [the 4th term in Eq.\ (\ref{eq:Replica})] 
in the wave-vector representation.
For the concise presentation, we introduce the operator 
{
$f^{\rm phy}_{{i}\tau{\bar \sigma}}\equiv{\tilde b}_{{i}\tau}{\tilde f}^{+}_{{i}\sigma}$. 
}
Then, the hybridization term in the original Hamiltonian [Eq.\ (\ref{H})] is transformed by the 
average over the random distribution of Pr ions to the following form: 
\begin{eqnarray}
& &
H^{\rm v}_{\rm hyb}=\sum_{\sigma=1}^N\sum_{\tau=1}^M\sum_{{\tilde i}}
V c^+_{{\tilde i}\tau\bar{\sigma}}f^{\rm phy}_{{\tilde i}\tau{\bar \sigma}}+{\rm h. c.}
,
\label{eq:A1}
\end{eqnarray}
{
where ${\tilde i}$ denotes the position on the {\it virtual} lattice.
By a usual prescription,
}
$f^{\rm phy}_{{\tilde i}\tau{\bar \sigma}}$ is expressed as 
\begin{eqnarray}
& &
f^{\rm phy}_{{\tilde i}\tau{\bar \sigma}}=
\frac{1}{\sqrt{N_{\rm f}}}\sum_{{\bf p}}e^{{\rm i}({\bf p}\cdot \mib{R}_{\tilde i})}
f^{\rm phy}_{{\bf p}\tau{\bar \sigma}}.
\label{eq:A2}
\end{eqnarray}
On the other hand, conduction electrons are defined on the original lattice points as noted below 
Eq.\ (\ref{eq:Replica}). Therefore, $c^{+}_{{\tilde i}\tau{\bar \sigma}}$ is given by $c^+_{{\bf p}\tau\bar{\sigma}}$ defined on 
all the original 
lattice points, i.e., other than the {\it virtual} lattice points, so that it is written as 
\begin{eqnarray}
& &
c^{+}_{{\tilde i}\tau{\bar \sigma}}=
\frac{1}{\sqrt{N_{\rm L}}}\sum_{{\bf p}}e^{-{\rm i}({\bf p}\cdot \mib{R}_{{\tilde i}})}
c^{+}_{\bf{p}\tau{\bar \sigma}}.
\label{eq:A3}
\end{eqnarray}
As a result, substituting expressions [Eqs. (\ref{eq:A2}) and (\ref{eq:A3})] into Eq.\ (\ref{eq:A1}), 
the hybridization in the {\it virtual} system is given by 
\begin{eqnarray}
& &
H^{\rm v}_{\rm hyb}=\sqrt{c_{\rm imp}}
\sum_{\sigma=1}^N\sum_{\tau=1}^M\sum_{{\bf p}}
Vc^{+}_{{\bf p}\tau{\bar \sigma}}f^{\rm phy}_{{\bf p}\tau{\bar \sigma}}+{\rm h. c.}, 
\label{eq:A4}
\end{eqnarray}
where $c_{\rm imp}=N_{\rm f}/N_{\rm L}$ is the concentration of Pr ions in the original lattice. 
Converting the wave vector representation into the real space one in the {\it virtual} lattice
{
with the use of inverse Fourier transformation of Eqs. (\ref{eq:A2}) and (\ref{eq:A3}),
}
the 4th term in Eq. (10) is recovered.
The $c_{\rm imp}$ dependence of Eq. (\ref{eq:A4}) is reasonable considering that the effect of localized Pr ions 
on the conduction electrons reflects the concentration of Pr ions, and  is crucial for understanding 
the relative importance of the impurities effect and the lattice effect in the single-channel Anderson 
model where the coefficient of $T^{2}$ term changes the sign from negative to positive at $c_{\rm imp}\sim0.5$ when 
the concentration of f-electrons increases from $c_{\rm imp}=0$ to $1$, i.e., 
from the local Fermi liquid to the heavy Fermi liquid behavior 
as discussed in Sect. 6. 
\section{Self-energy of conduction electrons in {\it virtual} system}
In this appendix, we show how the imaginary part of the self-energies 
[Eqs. (17)-(19)] is calculated.

{
\subsection{Self-energies of Fig. 3(a)}
}
First, we discuss how the expressions for the self-energies shown in Fig. 3(a) are 
obtained. To this end, we first rewrite Eq. (17) as
\begin{eqnarray}
      {\rm Im}\Sigma_{\rm c}^{(a)R}(\varepsilon=0;T)=-\tilde{C}
      +A_{\rm{imp}}T^2
      +2\tilde{C}\tilde{a}_f(T/\tilde{E}_0)^{\nu}\left(1-\frac{1}{M^2}\right),
      \non\\
      \label{b1}
\end{eqnarray}
where $\nu\equiv(1-\tilde{a}_f)\frac{M}{N}$.

The self-energy $\Sigma_{\rm c}^{(a1)}(i\varepsilon_n)$ shown by the Feynman diagram
of the first term in Fig. 3 (a) has no imaginary part, i.e.,
${\rm Im}\Sigma^{(a1)R}(\varepsilon=0)=0$ because $\Sigma^{(a1)}(i\varepsilon_n)$
is given as
\begin{eqnarray}
      \Sigma_{\rm c}^{(a1)R}(i\varepsilon_n)=\frac{\tilde{a}_fV^2}{M}\frac1{i\varepsilon_n-\tilde{E}_0},
      \label{b2}
\end{eqnarray}
so that
\begin{eqnarray}
      {\rm Im}\Sigma_{\rm c}^{(a1)R}(\varepsilon)
=-\pi\frac{\tilde{a}_fV^2}{M}\delta(\varepsilon-\tilde{E}_0).
      \label{b3}
\end{eqnarray}

The self-energy $\Sigma_{\rm c}^{(a2)}(i\varepsilon_n)$ shown by the Feynman diagram
of the second term in Fig. 3(a) gives the terms,
$
-\tilde{C}+A_{\rm{imp}}T^2$, in Eq. (\ref{b1})
as shown below.
The explicit form of the self-energy $\Sigma_{\rm c}^{(a2)}(i\varepsilon_n)$ is given by
\begin{eqnarray}
      \Sigma_{\rm c}^{(a2)}(i\varepsilon_n)
      &=&\tilde{V}^4T^2\sum_{\omega_m, \omega_l}\frac{1}{N_L}\sum_{{\bf p},\sigma}
      \bar{G}_{{\bf p}\tau\sigma}(-i\omega_m)
      \bar{F}_{i\sigma}(i\omega_l)\nonumber\\
&&      \times\bar{B}^2_{i\tau}(i\varepsilon_n+i\omega_l)
      \bar{F}_{i\sigma}(i\varepsilon_n+i\omega_l+i\omega_m)/\langle Q_i\rangle_\lambda,
      \label{b4}
\end{eqnarray}
where $\bar{G}_{{\bf p}\tau\sigma}(i\omega_n)$, $\bar{F}_{i\sigma}(i\omega_n)$,
and $\bar{B}_{i\tau}(i\nu_n)$
have been given by Eqs. (49), (51), and (15) in Ref. 10, respectively.
Substituting these expressions into Eq. (\ref{b4}),
$\Sigma^{(a2)}(i\varepsilon_n)$ is reduced to
\begin{eqnarray}
	&&      \Sigma_{\rm c}^{(a2)}(i\varepsilon_n)
       =\frac{\tilde{a}_f^2(T)\tilde{V}^4}{N}
       \int d\varepsilon \tilde{N}(\varepsilon)\non\\
& &
\qquad\qquad
       \times\left[
             \frac{f(-\varepsilon)}{(i\varepsilon_n-\tilde{E}_0(T))^2(i\varepsilon_n-\varepsilon)}
            -\frac{f(\varepsilon)}{(\varepsilon-\tilde{E}_0(T))^2(-i\varepsilon_n+\varepsilon)}
       \right],
       \non\\
      \label{b5}
\end{eqnarray}
where $\tilde{N}(\varepsilon)\equiv-(1/\pi N_f)\sum_{{\bf p},\tau}{\rm Im}\bar{G}_{\tilde{c}\tau\sigma}^{R}({\bf p}, \varepsilon)$, and $\tilde{E}_0(T)$ and $\tilde{a}_{\rm f}(T)$ are solutions of Eqs. (24) and (25) at finite temperature. 
Thus, after simple calculations, the imaginary part of $\Sigma_{\rm c}^{(a2)R}(\varepsilon=0; T)$ is given as
\begin{eqnarray}
      {\rm Im}\Sigma_{\rm c}^{(a2)R}(0; T)
      =-\frac{\pi}{N}\left(\frac{\tilde{a}_f(T)\tilde{V}^2}{\tilde{E}_0(T)}\right)^2\tilde{N}(0),
      \label{b6}
\end{eqnarray}
in the low temperature range $0\le T \ll E_0$.
At $T=0$, it is reduced to
\begin{eqnarray}
      {\rm Im}\Sigma_{\rm c}^{(a2)R}(0; 0)=-\tilde{C},
      \label{b7}
\end{eqnarray}
where
\begin{eqnarray}
      \tilde{C}\equiv\frac{\pi}{N}\left(\frac{\tilde{a}_f(0)\tilde{V}^2}{\tilde{E}_0(0)}\right)^2\tilde{N}(0),
      \label{b8}
\end{eqnarray}
which is nothing but Eq. (21).
The solutions of Eqs. (24) and (25) at finite temperature, $0<T\ll E_0$, change from those at $T=0$ as
\begin{eqnarray}
\tilde{E}_0(T)=\tilde{E}_0(0)+\Delta \tilde{E}_0(T),
      \label{b9}\\
\tilde{a}_f(T)=\tilde{a}_f(0)+\Delta \tilde{a}_f(T).
      \label{b10}
\end{eqnarray}
Using the Sommerfeld expansion in Eqs. (24) and (25), we obtain the modifications $\Delta \tilde{E}_0(T)$ and $\Delta \tilde{a}_f(T)$ in the lowest order in $T^2$ as
\begin{eqnarray}
      &&\frac{\Delta \tilde{E}_0(T)}{\tilde{E}_0(0)}\simeq\frac{\pi^2}{6}[1-\tilde{a}_f(0)]
      \left[\frac{T}{\tilde{E}_0(0)}\right]^2,
      \label{b11}\\
      &&\frac{\Delta \tilde{a}_f(T)}{\tilde{a}_f(0)}\simeq-\frac{\pi^2}{6}[1-\tilde{a}_f(0)][1+\tilde{a}_f(0)]
\left[\frac{T}{\tilde{E}_0(0)}\right]^2.
      \label{b12}
\end{eqnarray}
Substituting these results into Eq. (\ref{b6}), we obtain $A_{\rm imp}$ in Eq. (\ref{b1}) as
\begin{eqnarray}
      A_{\rm imp}&\equiv&
      \lim_{T\to0}\frac1{T^2}[{\rm Im}\Sigma^{(a2)R}(0; T)-{\rm Im}\Sigma^{(a2)R}(0; 0)]\nonumber\\
     &=&\frac{\pi^2}{3}\tilde{C}[1-\tilde{a}_f(0)][2+\tilde{a}_f(0)]\frac{1}{[\tilde{E}_0(0)]^2},
      \label{b13}
\end{eqnarray}
which is nothing but Eq. (27).

The self-energy $\Sigma^{(a3)}(i\varepsilon_n)$ given by the 
Feynman diagrams of the terms illustrated as dots in Fig. 3 (a), which is the third term in Eq. (\ref{b1}),
has the imaginary part 
\begin{eqnarray}
      {\rm Im}\Sigma_{\rm c}^{(a3)R}(\varepsilon=0; T)=2\tilde{C}\tilde{a}_f(0)[T/\tilde{E}_0(0)]^{\nu}(1-M^{-2}).
      \label{b14}
\end{eqnarray}
The explicit form of the dots in Fig. 3 (a) is shown in Fig. 4 in Ref.~\citen{Tsuruta2},
and that of ${\rm Im}\Sigma_{\rm c}^{(a3)R}(\varepsilon=0; 0)$ is given by Eq. (4.6) in Ref.~\citen{Tsuruta2}.

{
\subsection{Self-energy of Fig. 3(b)}
}
Second, we discuss why the self-energy $\Sigma_{\rm c}^{(b)}(i\varepsilon_n)$ shown by the Feynman diagram of Fig. 3 (b) is necessary~\cite{Miura}. 
The Green function of the conduction electrons of the order of $O[(1/N)^0]$ consists of series of terms including power series of $\Sigma_{ {\rm c}i}^{(a1)}(i\varepsilon_n)$ as
\begin{eqnarray}
      &&\bar{G}_{{\rm p}\tau\sigma}(i\varepsilon_n)\nonumber\\
      &=&
      \bar{G}^0_{{\rm p}\tau\sigma}(i\varepsilon_n)
      +[\bar{G}^0_{{\rm p}\tau\sigma}(i\varepsilon_n)]^2
      \frac1{N_f}\sum_i\Sigma^{(a1)}_i(i\varepsilon_n)\nonumber\\
      &&+[\bar{G}^0_{{\rm p}\tau\sigma}(i\varepsilon_n)]^3
      \frac1{N^2_f}\sum_{i\ne j}
      \Sigma^{(a1)}_i(i\varepsilon_n)\Sigma^{(a1)}_j(i\varepsilon_n)+\cdots,
      \label{b15}
\end{eqnarray}
where 
$\bar{G}^0_{{\rm p}\tau\sigma}(i\varepsilon_n)$ is the bare Green function of the conduction electrons.
The reason why only the terms $i\ne j$, etc., are taken into account is that the terms with $i=j$, etc., vanish after taking the limit $\{\lambda_i\}\to\infty$  in Eq. (4) for calculating expectation value of physical quantities.
Since the series of higher order terms of $\Sigma_i^{(a1)}(i\varepsilon_n)$ cannot be collected as a form of self-energy as it stands, we rearrange these terms as
\begin{eqnarray}
      &&\bar{G}_{{\rm p}\tau\sigma}(i\varepsilon_n)\nonumber\\
      &=&
      \bar{G}^0_{{\rm p}\tau\sigma}(i\varepsilon_n)
      \Bigg\{1
      +\bar{G}^0_{{\rm p}\tau\sigma}(i\varepsilon_n)
      \frac1{N_L}\sum_i\Sigma^{(a1)}_i(i\varepsilon_n)
      \non\\
	&&+\left[\bar{G}^0_{{\rm p}\tau\sigma}(i\varepsilon_n)
      \frac1{N_L}\sum_i\Sigma^{(a1)}_i(i\varepsilon_n)\right]^2+\cdots\Bigg\}\nonumber\\
      &&-[\bar{G}^0_{{\rm p}\tau\sigma}(i\varepsilon_n)]^2
      \left[\frac1{N^2_L}\sum_{i}\sum_{{\rm p}'}
      \bar{G}^0_{{\rm p}'\tau\sigma}(i\varepsilon_n)
\Sigma^{(a1)}_i(i\varepsilon_n)\Sigma^{(a1)}_i(i\varepsilon_n)+\cdots\right],
\non\\
      \label{b16}
\end{eqnarray}
where the first line can be collected as the self-energy 
$\frac1{N_L}\sum_i\Sigma_{ci}^{(a1)}(i\varepsilon_n)$.
The self-energy $\Sigma_{\rm c}^{(b)}(i\varepsilon_n)$ is defined by the expression in the bracket in the second line of Eq. (\ref{b16}),
and its explicit form
is given by
\begin{eqnarray}
      \Sigma_{\rm c}^{(b)}(i\varepsilon_n)
      &=&-\tilde{V}^4T^2\sum_{\omega_l, \omega_m}\frac{1}{N_L}\sum_{{\bf p},\sigma}
      \bar{G}_{{\bf p}\tau\sigma}(i\varepsilon_n)
      \bar{F}_{i\sigma}(i\omega_l)\bar{B}_{i\tau}(i\varepsilon_n+i\omega_l)
\nonumber\\
&&      \times\bar{F}_{i\sigma}(i\omega_m)\bar{B}_{i\tau}(i\varepsilon_n+i\omega_l)
      /\langle Q_i\rangle_\lambda^2.
      \label{b17}
\end{eqnarray}
Thus, after simple calculations, the imaginary part of $\Sigma_{\rm c}^{(b)R}(0; 0)$ is given as
\begin{eqnarray}
      {\rm Im}\Sigma_{\rm c}^{(b)R}(0; 0)=\tilde{C}\frac{1}{M^2},
      \label{b18}
\end{eqnarray}
which is nothing but Eq. (18).

\begin{figure}
      \includegraphics[width=10cm]{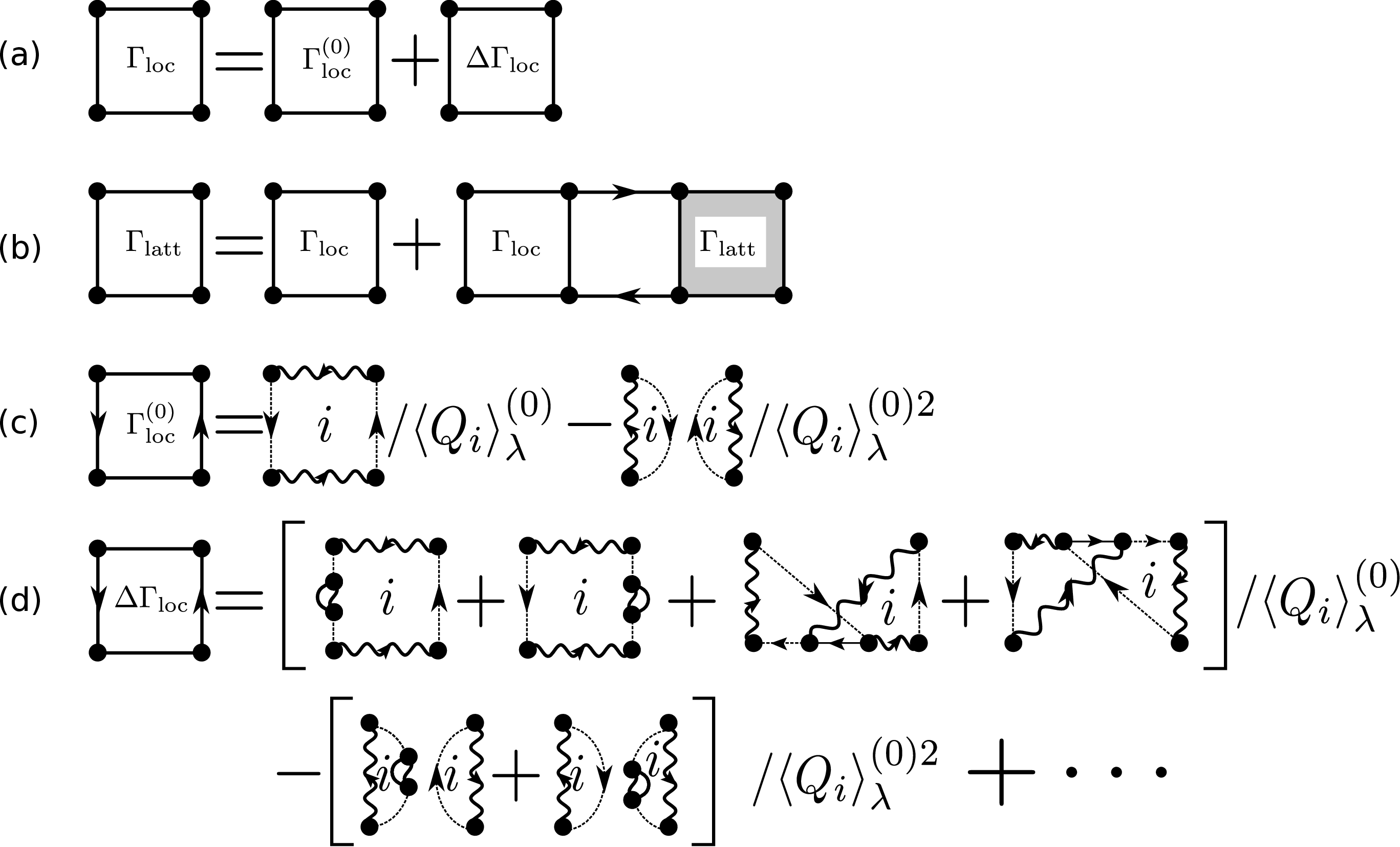}
      \caption{
            (a) Local vertex $\Gamma_{\rm loc}$, 
            (b) full vertex $\Gamma_{\rm q}$,
            (c) local vertex $\Gamma_{\rm loc}^{(0)}$ of $O[(1/N)^0]$,
            and (d) higher order corrections of local vertex $\Delta\Gamma_{\rm loc}$ in $1/N$-expansion.
            Notations are the same as those in Fig.\ \ref{Fig:SelfEnergy-1}.
}
      \label{figb1}
\end{figure}

{
\subsection{Self-energy of Fig. 3(c)}
}
Finally, we discuss the self-energy $\Sigma_{\rm c}^{(c)}(i\varepsilon_n)$ shown by the Feynman diagram of Fig. 3 (c).
The explicit diagrams included in vertices $\Gamma_{\rm loc}$ and $\Gamma$ in $\Sigma^{(c)}(i\varepsilon_n)$ are illustrated in Fig. \ref{figb1} (a) and (b), respectively.

If we use $\Gamma_{\rm loc}^{(0)}$ as the local vertex $\Gamma_{\rm loc}$ in Fig. \ref{figb1} (b),
the imaginary part of the self-energy, ${\rm Im}\Sigma_{\rm c}^{(c)R}(\varepsilon=0; T)$, is given by
      \begin{eqnarray}
            {\rm Im}\Sigma_{\rm c}^{(c)R}(\varepsilon=0; T)=
            \tilde{C}\frac{1}{1+\frac{\tilde{T}^{*(0)}}{T}}\frac{\tilde{T}^{*(0)}}{T}\left(1-\frac{1}{M^2}\right)-A_{\rm latt}T^2,
      \label{b19}
      \end{eqnarray}
as has been shown in discussions leading to
Eq. (42) in Ref. 10.
$T^{*(0)}$ in Eq. (\ref{b19}) is given by
\begin{eqnarray}
      T^{*(0)}=\frac{\tilde{E}_0}{1+\tilde{N}(0)\tilde{E}_0},
      \label{b20}
\end{eqnarray}
which appears in the factor $b$ of Eq. (59) in Ref. 10.

Following the method for deriving imaginary part of the self-energy of the Fermi liquid quasiparticles in Ref.~\citen{AGD2}, the coefficient $A_{\rm latt}$ is derived as
\begin{eqnarray}
	A_{\rm lattice}=\frac{\pi}{4}\tilde{C}^2N\frac{\tilde{a}_f(0)}{[\tilde{E}_0(0)]^3},
	\label{CImSigma3_A}
\end{eqnarray}
where we have approximated the full vertex $\Gamma(\vec{p}, \vec{p}_2; \vec{p}_1, \vec{p}+\vec{p}_2-\vec{p}_1)$ appearing in Eq. (19.31) of Ref.~\citen{AGD2} by $\Gamma^{ {\rm loc}(0A)}$, defined by Eq. (27) in Ref.~\citen{Tsuruta6} which is reduced to $2\tilde{a}_f^2\tilde{V}^4/M\tilde{E}_0^3$ in the static limit,
and $p_0v$ as $N\tilde{E}_0$ with $N=2$.
We have also assumed that the dispersion of the quasiparticles is given by that of the plane wave in three dimensions.
Note that the expression for the coefficient $A$ in Ref.~\citen{AGD2} is reduced to 
\begin{eqnarray}
	A=\frac14\left|\frac{a_f^2p_0}{\pi^2v}\Gamma\right|^2
\end{eqnarray}
if the wave vector dependence of the vertex $\Gamma$ is neglected.


The local vertex $\Delta\Gamma^{1a}_{\rm loc}$ shown by the Feynman diagram of the {\it first} term in the {\it first} bracket in Fig. \ref{figb1} (d) is given by 
\begin{eqnarray}
	&&\Delta\Gamma^{1a}_{\rm loc}(i\varepsilon_{n_1}, i\varepsilon_{n_2}; ix_l)
      =-\frac{1}{N_f}\sum_{\bf p}T^2\sum_{\nu_m,\omega_n}\bar{F}_{i\sigma}(i\nu_m)\bar{F}_{i\sigma}(i\nu_m+ix_l)
      \nonumber\\
	&&\quad \quad
      \times \bar{B}_{i\tau}(i\nu_n+i\varepsilon_{n_1})\bar{F}^2_{i\sigma}(i\nu_m+i\varepsilon_{n_2})\bar{G}_{\bf p\tau\sigma}(-i\omega_n)\non\\
	&&\quad \quad
	\times\bar{B}_{i\sigma}(i\nu_m+i\varepsilon_{n_2}+i\omega_n)/\langle Q_i\rangle_{\lambda_i},
      \label{b22}
\end{eqnarray}
where $\varepsilon_{n_i} (i=1, 2)$ are fermionic Matsubara frequencies and $x_l$ and $\nu_m$ are bosonic Matsubara frequencies, respectively.
After the analytic continuation from $i\varepsilon_{n_i}$ to $\varepsilon_i+i\delta$ and taking the limit $\varepsilon_1$ and $\varepsilon_2\ll T$, $\Delta\Gamma_{\rm loc}^{1a}$ [Eq. (\ref{b22})] is reduced to
\begin{eqnarray}
	\Delta\Gamma^{1a}_{\rm loc}(i\delta, i\delta; x_l)=-\frac1T\delta_{x, 0}\tilde{N}(0)\left(\frac{\tilde{a}_f\tilde{V}^2}{\tilde{E}_0}\right)^3\log\frac{T}{\tilde{E}_0}.
      \label{b23}
\end{eqnarray}
Refer to Appendix A in Ref.~\citen{Tsuruta2}  for an explicit calculation of the expression [Eq. (\ref{b22})].
Similarly, the singular terms in each diagram in Fig. \ref{figb1} (d) of the order of $O[(1/N)^n]$ is shown to be proportional to $[\log(T/\tilde{E}_0)]^n$ as can be seen in calculations performed in \S 4 of Ref.~\citen{Tsuruta2}.

In the case of single-channel, $M=1$, the singular terms cancel with each other,
while in the case of multichannel, $M\ge2$, the singular terms remain in the local vertex.
Just as ${\rm Im}\Sigma^{(a3)}$ includes singular factor $[T/\tilde{E}_0]^\nu$ in Eq. (\ref{b14}), the full local vertex $\Gamma^{(B)}_{\rm loc}$ illustrated in Fig. \ref{figb1} (d) includes the singular term proportional to $(T/\tilde{E}_0)^\nu$ as follows:
\begin{eqnarray}
	&&\Gamma^{(B)}_{{\rm loc}
}(i\delta,i\delta:ix_l)=\frac{\tilde{a}_{\rm f}^2\tilde{V}^4}{\tilde{E}_0^2}\frac{1}{T}\delta_{x_l,0}\left[1-2\tilde{a}_{\rm f}(T/\tilde{E}_0)^\nu\right]
\non\\
	&&\quad\quad
	\times\left(\delta_{\sigma_1,\sigma_4}\delta_{\sigma_2,\sigma_3}-M^{-1}\delta_{\sigma_1,\sigma_3}\delta_{\sigma_2,\sigma_4}\right),
      \label{b21}
\end{eqnarray}
which includes the higher order corrections in $(1/N)$ to $\Gamma^{(B)}$ in Eq. (29) of Ref.~\citen{Tsuruta6}.
(See also Ref.~\citen{Tsuruta5} for the dependence on $\sigma_1$, $\sigma_2$, $\sigma_3$, $\sigma_4$.)

An explicit form of the Bethe-Salpeter equation illustrated in Fig. \ref{figb1} (b) is given as
\begin{eqnarray}
	&&      \Gamma^{(B)}_{\bf q}(i\delta,i\delta:ix_l)=\Gamma^{(B)}_{\rm loc}(i\delta,i\delta:ix_l)\nonumber\\
	&&\quad\quad
	+\Gamma^{(B)}_{\rm loc}(i\delta,i\delta:ix_l)T\sum_{\varepsilon_n}\frac{1}{N_f}\sum_{\bf p}
      \bar{G}_{{\bf p}\tau\sigma}(i\varepsilon_n)
      \bar{G}_{{\bf p}+{\bf q}\tau\sigma}(i\varepsilon_n+ix_l)
      \non\\
	&&\quad\quad
      \times\Gamma^{(B)}_{\bf q}(i\delta,i\delta:ix_l).
      \label{b24}
\end{eqnarray}
By solving this equation, we obtain the full vertex $\Gamma^{(B)}_{\bf q}(i\delta,i\delta:ix_l)$ as follows:
\begin{eqnarray}
	&&\Gamma^{(B)}_{\bf q}(i\delta,i\delta:ix_l)=\frac{
		\frac{1}{\tilde{N}(0)}\frac{T^{*(0)}}T\left[1-2\tilde{a}_{\rm f}
      \left(\frac{T}{\tilde{E}_0}\right)^\nu\right]}
      {1+
	      \frac{T^{*(0)}}T\left[1-2\tilde{a}_{\rm f}\left(\frac{T}{\tilde{E}_0}\right)^\nu\right]}\non\\
	&&\quad\quad\times\left(\delta_{\sigma_1,\sigma_4}\delta_{\sigma_2,\sigma_3}-M^{-1}\delta_{\sigma_1,\sigma_3}\delta_{\sigma_2,\sigma_4}\right)\delta_{x_l, 0},
      \label{b25}
\end{eqnarray}
where the expression of the denominator is derived from the factor $K_{\rm q}(0)[K_{\rm q}(0)-T^{-1}f_{\rm q}^{(0,2)}(0)]$ in the denominator of Eq. (33) in Ref.~\citen{Tsuruta6}, and that of the numerator is given by $\Gamma_{\rm loc}$ [Eq. (\ref{b21})].
By calculating the self-energy in Fig. 3(c),
we obtain the imaginary part of the self-energy $\Sigma_{\rm c}^{(c)R}(\varepsilon=0; T)$ as
      \begin{eqnarray}
            {\rm Im}\Sigma_{\rm c}^{(c)R}(\varepsilon=0; T)=
            \tilde{C}\frac{1-2\tilde{a}_{\rm f}(T/\tilde{E}_0)^\nu}{1+\frac{\tilde{T}^*}{T}}\frac{\tilde{T}^*}{T}\left(1-\frac{1}{M^2}\right)-A_{\rm latt}T^2,
	    \non\\
      \label{b26}
      \end{eqnarray}
      where $\tilde{T}^*$ is given by 
      \begin{eqnarray}
	      \tilde{T}^*=\tilde{T}^{*(0)}[1-2\tilde{a}_f(T/\tilde{E}_0)^\nu].
	      \label{b27}
      \end{eqnarray}
      Equation (\ref{b26}) is nothing but Eq. (19).

\begin{thebibliography}{99}
\bibitem{Comment_nu}
This expression for $\nu$ is essentially the same as Eq. (4.6) in 
Ref.~\citen{Tsuruta2} in which 
$M\ll N$ has been assumed. However, the exact expression in the case of $M=N$ as in the 
present case, is expected to be  $\nu\equiv (1-{\tilde a}_{\rm f})M/(N+M)$ because 
this exponent was derived by the NCA (Non-Crossing Approximation) 
that is considered to be valid at the high-$T$ 
region, $T>{\tilde E}_{0}(0)\sim T_{\rm K}$, as discussed in Ref.~\citen{Cox3}. 
Namely, in that case, the relation $\nu=1/2$ holds when $N=M=2$, which is consistent with the 
correct result in the two-channel Kondo model given in Ref.~\citen{Ludvig}.
\bibitem{Comment0}
In Ref.~\citen{Tsuruta6}, the factor $[1-2{\tilde a}_{\rm f}(T/{\tilde E}_{0})^{\nu}]$ in 
Eq.\ (\ref{ImSigmaTotal}) has been approximated as 1 because they were interested in the $T$ region, 
$T<{\tilde T}^{*}\sim{\tilde E}_{0}(0)\sim T_{\rm K}$, where the factor 
$2{\tilde a}_{\rm f}(T/{\tilde E}_{0})^{\nu}$ gives only minor correction to the $T$ dependence 
of ${\rm Im}\Sigma^{\rm R}_{\rm c}(\varepsilon=0,T)$.
\end{thebibliography}
\end{document}